\documentclass[aps,prl,twocolumn,showpacs,superscriptaddress]{revtex4-1}

\usepackage{graphicx}
\usepackage{amsmath}
\usepackage{verbatim}
\usepackage{hyperref}

\begin{document}
\vfuzz2pt 
\hfuzz2pt 

\title{Cavity QED in a molecular ion trap}
\date{\today}

\author{D. I. Schuster}
\affiliation{Departments of
Applied Physics and Physics, Yale University, New Haven, CT 06520}
\author{Lev S. Bishop}
\affiliation{Department of Physics, Yale University, New Haven, CT 06520}
\author{I. L. Chuang}
\affiliation{Center for Ultracold Atoms, Massachusetts Institute of Technology, Cambridge, MA 02139}
\author{D. DeMille}
\affiliation{Department of Physics, Yale University, New Haven, CT 06520}
\author{R. J. Schoelkopf}
\affiliation{Departments of
Applied Physics and Physics, Yale University, New Haven, CT 06520}

\begin{abstract}
We propose a class of experiments using rotational states of dipolar molecular ions trapped near an on-chip superconducting microwave cavity.  Molecular ions have several advantages over neutral molecules for such cavity quantum electrodynamic (QED) experiments.  In particular, ions can be loaded easily into deep RF traps, and are held independent of their internal state.  An analysis of the detection efficiency for, and coherence properties of, the molecular ions is presented.  We discuss approaches for manipulating quantum information and performing high resolution rotational spectroscopy using this system.  
\end{abstract}


\maketitle

\section{Introduction}
Studies of ultracold molecules present opportunities to test fundamental laws of physics~\cite{Sauer2005,Sinclair05,Schiller05,demille043202,zelevinsky043201,demille023003,muller_tests_2004}, manipulate quantum information~\cite{demille_quantum_2002,andre_coherent_2006,rabl_hybrid_2006}, and better understand low temperature chemistry~\cite{balakrishnan_chemistry_2001}.  Trapping, cooling, and manipulating molecules has thus been a long-standing goal.  Progress has been difficult because excited electronic states decay to a number of internal vibrational and rotational states, rendering laser cooling ineffective in most molecules (although recent progress in this direction is promising~\cite{Shuman09}).  Despite this difficulty there has been substantial success recently both with clouds of neutral molecules and with clouds of molecular ions created using a variety of methods; see e.g.~\cite{Ni09182008,bethlem_electrostatic_2000,weinstein_magnetic_1998,molhave_formation_2000,Roth05} and for reviews see e.g.~\cite{KremsColdMoleculesReview,Carr09}.

One interesting application of ultracold molecules is as a component of a hybrid quantum information processing system.  One class of hybrid architecture employs a high finesse cavity as an interface to a natural quantum system~\cite{tian_interfacing_2004,tian_coupled_2004,verdu_strong_2009,imamoglu_cavity_2009}.  In particular a recent proposal described ideas for employing an all-electrical interface between rotational states of neutral polar molecules and superconducting cavities~\cite{andre_coherent_2006}.  Proposals with neutral molecules have also suggested using thermal clouds~\cite{rabl_hybrid_2006} or self-assembled dipolar crystals~\cite{rabl_molecular_2007} as a quantum memory for solid-state qubits.  If successfully implemented, this system would represent a novel pathway for control of long-lived rotational degrees of freedom and a significant step forward in quantum information processing.  

To date, there have not been any experimental realizations of these promising proposals, partially because of the difficulty in trapping sufficient densities of neutral molecules and detecting their rotational state.  Many of these challenges arise from the DC Stark trap generally used to hold neutral molecules \cite{bethlem_electrostatic_2000,Meek2009}, a type of trap that is based on the shift of the rotational states due to the trapping fields.  Such traps only allow depths less than the rotational transition energy [$\sim\! 1\,\rm{K}\,(\approx\! 20\,\rm{GHz})$ for species of interest here]; moreover, for hybrid quantum devices the molecules must be held at high density ($n \sim 10^{12}$ cm$^{-3}$). Hence, even for initial experiments the proposed schemes using neutral molecule-based hybrid devices require a molecular phase-space density that is difficult to reach--particularly taking into account the need to integrate the molecule trap with a cryogenic apparatus suitable for superconducting cavities.  Further, the rotational ground state cannot be confined in such a DC Stark trap; hence relaxation causes not only loss of information, but also loss of molecules.  Finally, because shifts of the rotational states are used for trapping, dephasing of rotational superpositions is typically severe (with dephasing rate comparable to the trap depth); this can be mitigated sufficiently only if the molecules are ultimately cooled to extremely low temperatures \cite{andre_coherent_2006}.  

Here we present an alternative approach, employing molecular ions (rather than neutral molecules) held inside a superconducting cavity integrated with an ion trap.  We show that ions address several of the obstacles present in realizing neutral molecule-based experiments of this type.  Molecular ions, can use a planar surface electrode Paul trap \cite{chiaverini_surface-electrode_2005,seidelin_microfabricated_2006} integrated into the microwave cavity. Such planar traps typically have depths of $\sim\!1000\,\rm{K}\,(\approx\! 0.1\, \rm{eV})$ and are fabricated lithographically, allowing micron-scale traps to be realized.  This trapping is independent of the internal state of the molecular ion and, with appropriate cooling, should hold the molecules for hours or longer.  We find that molecular ions can have much longer internal state coherence times than neutral molecules at similar, experimentally realizable  temperatures.  Moreover, ion crystals containing many thousand molecular ions have been realized \cite{drewsen_large_1998}, making ensemble-based schemes potentially viable.  Just as for neutral species, here thermal motion causes the molecules to experience different electric fields inside the trap, leading to the dephasing of rotational-state superpositions.  However, in an ion trap, the ion's charge guarantees that the molecule is trapped at a position where the net average electric field is zero.  For all molecules, the DC field null corresponds to a ``clock point'' or ``sweet spot'' \cite{vion_manipulatingquantum_2002}, where the spittings between rotational states are sensitive to electric field fluctuations only at second order \cite{andre_coherent_2006}.  This substantially reduces the effect of several possible dephasing mechanisms.  

Due to these properties, molecular ions could present a practical route for realizing an all-electrical coupling to polar molecules, with current technology.  However, with these advantages also come new challenges due to e.g. space-charge effects and the use of large RF trapping fields.  As we discuss below, such effects ensure that even with molecular ions, cavity QED experiments will remain technically challenging.  Nevertheless, we believe this type of experiment to be sufficiently interesting that it warrants the investigation here.

We analyze three classes of experiments.  In one, the aim is to perform rotational spectroscopy on a cloud of molecules in a cryogenic buffer gas. This is motivated by the fact that despite their importance in many areas of chemistry and physics~\cite{Wing1988,Asvany08,Koelemeij07}, there is a relative paucity of spectroscopic data on molecular ions~\cite{Herzbergconstants,Saykally81,Hirota92}  (as opposed to neutral molecules), especially for species other than hydrides. Our proposed methods would provide a new way to determine rotational constants and hyperfine structure of many species.  The other two types of proposed experiments, to be performed in ultrahigh vacuum conditions, focus on the manipulation of quantum information.  In one, we use a single molecule as a qubit; in the other, a single chain of molecular ions is used as a quantum memory.   Our goal for the qubit experiment is to reach the strong coupling limit of cavity QED~\cite{WallsMilburn}.  A quantum memory relaxes the requirement for single-molecule strong coupling but is not sufficient by itself for quantum computing.  Instead a large number of molecules which collectively achieve this limit are used to store quantum information from a solid-state qubit~\cite{rabl_hybrid_2006}. There are two requirements for a useful memory: the coupling should be strong enough to transfer a state in less than a qubit lifetime and simultaneously the decoherence rate of the memory must be significantly less than that of the qubit.   For each of these experiments, we discuss the strength of the molecule-cavity coupling, sources and rates of decoherence, and the time required to measure the relevant parameters of the system.

\section {Description of experimental setup}

\begin{figure}
\includegraphics{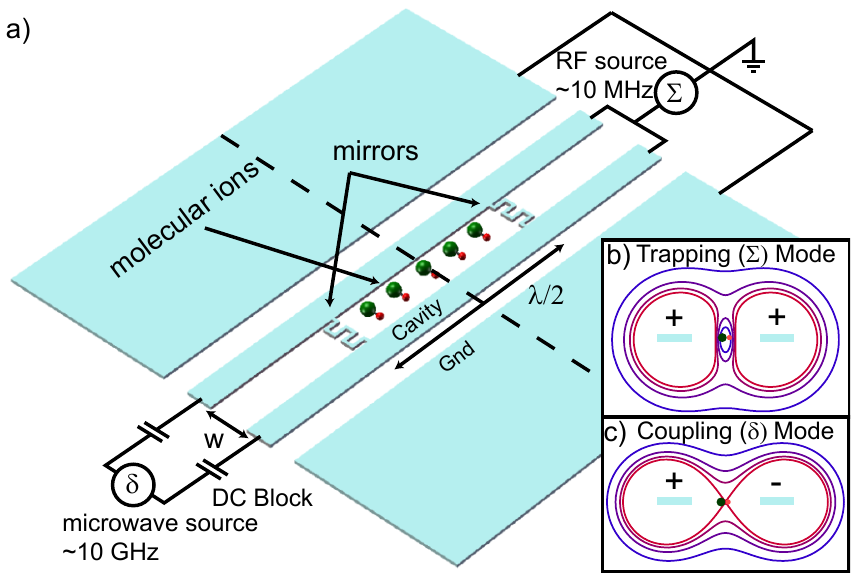}
\caption{\textbf{a)} Schematic of molecular ions trapped inside an integrated microwave resonator and split-ground ion trap.  The inset shows contours of constant electric field magnitude ($\left|\mathcal{E}\right|$) in response to voltages on the two center electrodes.  The two center electrodes are used both for trapping and for coupling molecules to the cavity.  Electric field contours are colored from low (blue) to high (red) $\left|\mathcal{E}\right|$. \textbf{b)} The trapping ($\Sigma$) mode, when driven at $\Omega_{\rm{RF}}$, forms a dynamically stable minimum between the two electrodes.  \textbf{c)} The coupling ($\delta$) mode is coplanar-stripline with inductive shorts forming a half-wave ($\lambda/2$) microwave cavity at the molecular rotation frequency ($\omega_{01}\sim 10\, \rm{GHz}$).  At the trap location, the $\delta$ mode has a saddle point in $\left|\mathcal{E}\right|$, with a large electric field for coupling to the molecules' rotational dipole. \label{fig:schematic}}
\end{figure}

Our experiments require dipolar molecular ions to be trapped in close proximity to a high finesse cavity.  Two-dimensional surface electrode Paul traps have been demonstrated in a variety of geometries and sizes~\cite{chiaverini_surface-electrode_2005} and have recently been made from superconducting thin films which have also been used for fabricating suitable microwave transmission line cavities~\cite{day_broadband_2003}.  In order to have coupling between the cavity and the molecules, they should be suspended close to an anti-node of the microwave electric field.  Though there are several viable geometries~\cite{chiaverini_surface-electrode_2005}, one possibility is shown in Fig.~\ref{fig:schematic}.  Trapping is achieved by applying an even mode ($\Sigma$, Fig.~\ref{fig:schematic}b) RF voltage to the two center electrodes.  At the same time, a microwave frequency voltage on the odd mode ($\delta$, Fig.~\ref{fig:schematic}c) of the center electrodes couples to the rotational dipole.  A high finesse microwave cavity is formed using inductive shorts as mirrors, which set the wavelength $\lambda$ and resonance frequency $\omega_{\rm{r}}$.  The cavity decay rate $\kappa$ and quality factor $Q=\omega_{\rm{r}}/\kappa$ are set by the inductance of the shorts.  This geometry yields an efficient coupling and should be scalable in width $w$ from millimeters to microns.

We consider diatomic species where the rotational states have energies $E_{J} = \hbar B J (J+1)$, where $B$ is the molecular rotation constant and $J$ is the angular momentum quantum number, giving transition frequency $\omega_{01} = 2B$ between the $J=0$ and $J=1$ rotational states.  For concreteness we take CaCl$^+$ as an example molecular ion, chosen for its convenient $B/2\pi=4.5\, \rm{GHz}$ and its large dipole moment $\mu \sim 12\, \rm{Debye}$~\cite{Raouafi1999248,Kotochigova2008}, corresponding to a transition dipole moment $p=\mu/\sqrt{3}\simeq 3\, e a_0$, where $a_0$ is the Bohr radius.  While CaCl$^+$ is a promising candidate species, in principle any molecule with $B/2\pi \approx 1-20\, {\rm GHz}$ and $\mu >1 \, {\rm Debye}$ should be feasible, enabling the use of a large class of molecules.  

The strength of the cavity-molecule coupling is critically important in all three classes of experiment, with larger values always preferable.   This strength is best characterized by the rate at which a photon is coherently absorbed and emitted by a molecule inside the cavity, known as the vacuum Rabi rate, $g$.  For a single molecule in a resonator with wave impedance $Z_0$ and width $w$, this is given by $g=g_1 \equiv \beta[p/(e w)][Z_0/(2R_q)]^{1/2}\omega_{01}$ \cite{schoelkopf_wiring_2008}. Here $\beta$ is the ratio of the electric field at the molecule's location to the maximum field in the resonator and $R_q=\hbar/e^2$ is the resistance quantum.  Surface electrode traps have been made as small as $w\approx 40\mu$m limited primarily by optical access~\cite{seidelin_microfabricated_2006}.  The minimum theoretical size of an ion trap has been estimated at $w_{\rm min}\sim 100\,$nm~\cite{chiaverini_surface-electrode_2005}.   We discuss traps with $w \sim 1000 - 10~\mu$m, where $\beta \approx 1/2$ is typical and hence $g_1/2\pi \sim 65\,\rm{Hz}-6.5\,\rm{kHz}$. The coupling $g_N$ to an ensemble of $N$ molecules in the cavity is substantially enhanced compared to the single-ion coupling: here for molecules near the center of the resonator $g = g_N \equiv g_1 \sqrt{N}$  \cite{tavis_exact_1968}.  Increasing $w$ at constant ion density $n_{\rm i}$ does not alter the collective coupling because the gain in number of ions is canceled by the reduction in $g_1$.  The density $n_{\rm i}$ is established when the Coulomb repulsion of the ions is comparable to the trapping force.  Under typical conditions $n_{\rm{i}} \approx (10 ~\rm{\mu m})^{-3}$.

In addition to the coherent coupling between the molecules and cavity, it is also important to consider the coupling to the continuum, which determines the ability to initialize and extract the state of the molecules.  This interaction is quantified by the spontaneous emission rate of the molecules' rotation.  In free space this relaxation rate is inconveniently slow, $\Gamma_{\rm free}/2\pi \approx 10^{-7}\, {\rm Hz}$.  In the one-dimensional environment of a transmission line, the rate $\Gamma_{\rm tx}=g^2/\omega_{01}$ can be enhanced by a factor of $\lambda^2/w^2\approx 10^2-10^6$.  It can be further enhanced by placing it in a near-resonant cavity giving $\Gamma_{\kappa}\approx g^2/\kappa$ (with a maximum rate of $\kappa/2$). By dynamically detuning the cavity frequency~\cite{sandberg_tuningfield_2008} from the rotational transition of the molecules, emission can also be suppressed allowing more time for coherent operations.  

The ion trap design has a significant influence on the coupling strength of molecules to the cavity, as well as on the decoherence properties of superpositions of rotational states of the molecules.    In order to understand the scaling of both coupling strength and decoherence rates with trap parameters, it is useful to review ion trap dynamics.  An ion with charge $Q_{\rm{i}}$ and mass $m$ is trapped at the node of an AC electric quadrupole field created by a voltage $V_{\rm{RF}}$, applied at drive frequency $\Omega_{\rm{RF}}$.  The trap is dynamically stable (in the absence of additional DC fields) when the Mathieu $q$ parameter
\begin{equation}
q = \frac{Q_{\rm{i}} V_{\rm{RF}}}{\frac{1}{2} m r_0^2 \Omega_{\rm{RF}}^2}
\end{equation}
satisfies $q\leq0.5$, where $r_0\sim2 w$ is the effective radius of the trap.  This requirement links the drive voltage (and thus trap depth) and drive frequency to the size of the trap.  We show below that this linkage leads to increased decoherence rates for rotational superpositions, as one scales to traps with smaller values of $w$ to get stronger couplings $g$.  When the stability condition is satisfied, the particles feel an effective harmonic potential~\cite{paul_electromagnetic_1990}, with secular trap frequency $\omega_{\rm{t}} = 2^{-3/2} q \Omega_{\rm{RF}}$, and potential energy $U=(r/r_0)^2 U_{\rm{max}}\propto\left|\mathcal{E}\right|^2$, up to a maximum depth $U_{\rm{max}}=\eta_{\rm{t}} q Q_{\rm{i}} V_{\rm{RF}}/8$, where $E$ is the RF electric field and $\eta_{\rm{t}}$ is the trap efficiency.  Typically $\eta_{\rm{t}} \sim 10\%$ and $r_0\sim 2 w$ for a planar trap~\cite{chiaverini_surface-electrode_2005}, and we take $q=0.2$ for our estimates.

\section{Decoherence}

Decoherence of rotational superpositions, especially dephasing, arises when electric fields varying in space or time lead to uncontrolled Stark shifts.  Since the ions are trapped, the net time averaged field they experience is identically zero.  However, any AC field still gives the rotational transitions a second order Stark shift $\hbar \delta\omega_{01}(\left|\mathcal{E}_{\rm rms}\right|^2) \approx p^2 \left|\mathcal{E}_{\rm rms}\right|^2/\hbar \omega_{01}$.  In a Paul trap, any movement of the molecular ion away from the null of the rms RF electric field (i.e., from the minimum of the effective trap potential) exposes the molecule to large oscillating fields from the trap itself.  If the magnitude of these fields is not reproducible, they lead to decoherence.  Three effects are primarily responsible for such spatial deviations of the ions: thermal motion within the trap, stray DC electric fields, and space charge (i.e. Coulomb forces from neighboring ions).  

\subsection{Dephasing due to thermal motion and stray electric fields}

Effects due to thermal motion and stray DC fields can be estimated as follows.  If the DC potentials on the trap leads are small, the time-averaged rms electric field felt by a single trapped ion can be written as \cite{berkeland_minimization_1998-1} 
\begin{equation}\label{eq:Erms}
\left|\mathcal{E}_{\rm rms}\right|^2 = \left|\mathcal{E}_{\rm th}\right|^2 + \left|\mathcal{E}_{\rm off}\right|^2, 
\end{equation}
where
\begin{equation}\label{eq:Eth}
\left|\mathcal{E}_{\rm th}\right|^2 \approx \frac{m \Omega_{\rm RF}^2}{Q_{\rm i}^2}k_{\rm B} T _m\end{equation}and
\begin{equation}\label{eq:Erm}
\left|\mathcal{E}_{\rm off}\right|^2 \approx 8 \left(\frac{\mathcal{E}_{\rm DC}}{q}\right)^2.
\end{equation}
Here $T_m$ is the motional temperature and $\mathcal{E}_{\rm DC}$ is the stray DC electric field due to uncompensated charges on the trap electrodes.  The first term, $\left|\mathcal{E}_{\rm th}\right|^2 $, arises from thermal motion exposing the ion to the RF trapping fields.  We find it useful to define a dimensionless parameter
\begin{equation}
\Lambda = \frac{\delta\omega_{01}(\left|\mathcal{E}_{\rm th}\right|^2)}{k_bT_m} \simeq \frac{8}{q^2}\left(\frac{p}{e x_0}\right)^2\frac{\omega_{\rm{t}}}{\omega_{01}},
\end{equation}
which is a measure of the sensitivity of the rotational frequency to a finite motional temperature of the ions.  The thermal dephasing rate can thus be written as \begin{equation}\label{eq:thermaldephasing}
\Gamma_{\rm th}= \Lambda k_{\rm B} T_m/\hbar.
\end{equation}
The ``suppression factor'' $\Lambda$ depends on the ratio of the rotational dipole, $p$, to the motional dipole, $e x_0$, where $x_0=(\hbar/m \omega_{\rm{t}})^{1/2}$ is the zero-point motion of the molecule in the trap; and to the ratio of the trap motional frequency to the rotational frequency.  Thus at fixed stability ($q$), low frequency traps (small $\omega_{\rm t}$) using light molecules (large $x_0$ and $\omega_{01}$), are least dephasing (see Table~\ref{tab:parameters}).  

Thermal dephasing in any given geometry can be minimized by cooling the motion to the lowest possible temperature.  We emphasize that for similar motional temperatures, the thermal dephasing rate for molecular ions is far smaller---by roughly the factor $\Lambda$---than for neutral molecules held in typical DC Stark traps (see Ref. \cite{andre_coherent_2006}).  This is due in part to the automatic ``sweet spot'' biasing for the ions, and in part to the large restoring force on the ion's charge, which prevents the ion from sampling large AC electric fields simply due to random thermal motion.  When compared to neutrals held in an optimally-biased DC Stark trap \cite{andre_coherent_2006}, the advantage of ions is reduced; however, even here thermal decoherence rates for ions are smaller than for neutrals held in traps of similar strength and scale, as long as the motional temperature satisfies $T_m \gtrsim \hbar \omega_t /k_{\rm B}$, or when the number of thermal motional quanta $\bar{n}$ satisfies $\bar{n} \gtrsim 1$.   

The second term in Eq.~\ref{eq:Erms} arises from DC electric fields but should not be confused with the true DC Stark shift felt by neutral molecules, as for a trapped ion the time averaged electric field is always zero.  Rather, the stray field causes the ion to be displaced from the trap center, exposing it to larger RF trapping fields.  Because of the quadratic sensitivity of the Stark shift $\delta\omega_{01}$ on $\left| E_{\rm rms}\right|$, the overall effect of the stray DC field is hence a small net AC Stark shift.  In fact, the offset-induced rms field $\left| E_{\rm off}\right|$ is typically larger than the stray field, with $\left|\mathcal{E}_{\rm off}\right| \sim 10 \times \left|\mathcal{E}_{\rm DC}\right|$.  This leads to a shift of $\Gamma_{\rm st}/2\pi \sim 10\left|\mathcal{E}_{\rm DC}\right|^2/(1\, {\rm V}^2{\rm /m}^2)\,\rm{Hz}$~\cite{berkeland_minimization_1998-1}.   The dephasing times of superconducting qubits~\cite{schuster_ac_2005} imply a voltage noise on the center electrode equivalent to $\left|\mathcal{E}_{\rm DC}^{\rm min} \right| \lesssim 10\, \mu\mathrm{V}/w$, giving dephasing rates of only a few Hz (see Tab.~\ref{tab:parameters}).  

\subsection{Dephasing due to space charge effects}

When an ensemble of ions, rather than a single ion, is in the trap, space charge (i.e., Coulomb forces between the ions) can prevent all of the ions from sitting at the field null, exposing some ions to spatial inhomogeneity of the trapping fields.  A simple model of a cylindrical cloud of ions, with density determined by force balance between space charge and the trapping force, yields a maximum Stark shift of $\hbar \delta \omega_{01} = 4 \Lambda U_{\rm{max}} N/\left( n_{\rm{i}} \pi w^2\lambda \right)$; we take $\Gamma_{\rm{sc}} = \delta \omega_{01}$ as the space-charge induced dephasing rate.  

For the spectroscopy experiment, we consider ion clouds with the largest possible number of ions $N$, where detection signals are maximized but $\Gamma_{\rm{sc}}$ is substantial.  For the quantum memory experiment, it is necessary to simultaneously maximize $N$ and minimize $\Gamma_{\rm{sc}}$.  This can be achieved by limiting $N$ such that the ion cloud forms a linear chain of individual ions along the length of the resonator.  (This is possible because the transverse confining potential of the trap is much steeper than the longitudinal potential, which is essentially flat over most of the length of a linear ion trap.)  In this regime, the dephasing due to space charge is eliminated; hence we ignore $\Gamma_{\rm{sc}}$ for the memory experiment as well as for the qubit (single ion) case.  

We note in passing that at sufficiently low temperatures, the ion cloud can crystallize.  In this regime, the positions of the ions are fixed, with each position experiencing a different time-independent Stark shift.  Since the shifts are coherent, the ensemble could be refocused using spin echo techniques.  If rapid diffusion of ions in cold clouds or crystals is taken into account, the Stark shifts can be motionally narrowed, reducing dephasing.  However, in our estimates of space-charge induced dephasing, we ignore these potentially large advantages and instead quote decoherence rates as deduced from the total rms electric field experienced in the ensemble.

\subsection{Dephasing and relaxation due to collisions}

Molecules can also decay or dephase through collisions, either with other molecules or with a deliberately introduced buffer gas.  Collisions with buffer gas thermalize the motional and rotational state of the molecules and can be useful in many ways. For example, buffer gas can be used to cool molecular ions in-situ as a means to load the trap.   Above $T_c \sim 1.5\, {\rm K}$ $^4$He or $^3$He gas will have sufficient density that it can be used to relax the rotational states at a useful rate $\Gamma_{\rm bg}$, reaching equilibrium when the rotational temperature $T_r \approx T_c$ \cite{weinstein_magnetic_1998}.  Our proposed spectroscopy experiment operates in this regime.  However, the qubit and memory experiments require removing any buffer gas, and using additional motional and rotational cooling by some alternate method \cite{andre_coherent_2006}.  

Even in the absence of buffer gas, the molecules can collide with one another, causing decay or dephasing depending on the duration and distance of closest approach of the collision.  To estimate the effects of collisions we assume worst case head-on collisions between ions with no screening.  Ions are assumed to have thermal velocity, $v_{\rm th} = (3 k_{\rm B} T_m/2 m_i)^{1/2}$, and the distance of closest approach is determined when this kinetic energy is fully converted into Coloumb repulsion potential, giving a distance of closest approach $r_{\rm min} = Q_i^2/6 \pi \epsilon_0 k_{\rm B} T_m$, and duration $\delta t = 2 r_{\rm min}/v$.  Depending on their duration relative to the rotational period collisions can cause relaxation and/or dephasing.  At $T_m=1000\,$K the collision time is $\delta t \sim100\,$ps, and significant relaxation will occur.  Below $T_m=10\,$K, where the proposed experiments would be performed, the collision duration is $\delta t \sim 100\,$ns, resulting in negligible relaxation.  This is in contrast to neutral-neutral collisions~\cite{rabl_hybrid_2006} which have a much shorter range interaction.  Below $T_m=10\,$K each collision gives the rotation a phase kick $\delta \phi =\delta \omega \delta t$, with $\delta \omega = p^2 (Q_i/4 \pi \epsilon_0 r_{\rm min}^2)^2/\hbar \omega_{01}$ the maximum Stark shift during the collision.  At these temperatures the phase kick is small and the dephasing is diffusive or ``motionally narrowed'' giving a collisional dephasing rate $\Gamma_{\rm c}=\delta \omega^2 \tau$, where $\tau = 1/n_i v_{\rm th} \sigma$ is the time between collisions.  The molecule cross-section is given by $\sigma=4\pi r_{\rm min}^2$ and the density is assumed to be limited by space charge, $n_i = 3 m_i q^2 \epsilon_0 \Omega_{\rm rf}^2/2Q^2$.  Once again we note that at sufficiently low temperatures the ensemble will crystallize, stopping collisions entirely; however again we do not include this effect in our estimates.  Compared with ensembles of neutral molecules, ionic ensembles are thus very robust against decoherence due to collisions.  This makes the linear chain of molecular ions potentially attractive for use as a quantum memory in a hybrid atomic/solid-state device of the type described in Ref.~\cite{rabl_hybrid_2006}.  

\subsection{Hyperfine encoding}

For the qubit and quantum memory experiments, the total decoherence rate of molecular superpositions is among the most crucial parameters.  Much as for neutral molecules~\cite{andre_coherent_2006}, these rates can be dramatically suppressed by encoding the quantum information into superpositions of hyperfine structure (HFS) sublevels of a single rotational state.  Such HFS-encoded states are dephased by fluctuating electric fields at rates a factor $\Upsilon \sim \omega_{\rm HFS}/\omega_{01}$ smaller than the rates for rotational superpositions.   Here, $\hbar \omega_{\rm HFS}$ is the HFS splitting in a single rotational state.  For molecules such as CaCl$^+$, in a $^1\Sigma$ electronic state and with only one non-zero nuclear spin $I\ge 1$ (e.g. $I({\rm ^{35}Cl}) = 3/2$), the qualitative features of the HFS are as follows.  HFS does not affect the rotational state $J=0$, but splits the $J=1$ state into three sublevels (with total angular momentum $F=I-1,I,I+1$) due to the interaction of the nuclear electric quadrupole moment with the internal electric field gradient of the molecular electrons.  Although $\omega_{\rm HFS}$  is not known for CaCl$^+$, for neutral molecular species with similar structure (e.g. KCl or RbCl), typical values are $\omega_{\rm HFS}/2 \pi \sim 0.1 - 1$ MHz \cite{Cederberg06,Nitz84}.  Hence we estimate $\Upsilon \sim 10^{-4}$.  We note in passing that even greater suppression of decoherence might be possible by encoding information into Zeeman sublevels of the HFS sublevels, since here levels with angular momentum projection quantum numbers $+m_F$ and $-m_F$ are exactly degenerate on application of electric fields.  However, we do not explore this possibility further here.

\section{Detection}
Using the coupling strength and decoherence rates, we calculate the signal-to-noise for detection of the system as follows.  The measurement is performed by monitoring the transmission of a microwave signal through a resonator which couples to the molecules, and comparing it to a reference signal which bypasses the cryostat.  The signal strength is primarily set by the amount of power that can be absorbed by the cloud without saturating, while the noise is ideally determined by the microwave amplifier.  For a good cryogenic amplifier the noise temperature is $T_N \sim 5$~K\cite{Cryogenic1988}, corresponding to an effective noise photon number $n_{\rm amp} = k_{\rm B} T_N /\hbar \omega_{01} \sim 20$.  The detection rate, $\Gamma_{\rm{m}}$ will be defined as the inverse of the time required to attain a signal to noise ratio of unity.

To facilitate this discussion we define the cooperativity $C=2 M_0 g^2/\kappa \Gamma_2$, a dimensionless figure of merit.  Here $M_0=N \tanh \left(\hbar \omega_{01} / k_{\rm{B}} T_r\right)$ is the number of participating molecules at rotational temperature $T_r$, and $\Gamma_2=\Gamma_1/2+\Gamma^{\rm rot}_{\phi}$, where $\Gamma_1 = \Gamma_\kappa + \Gamma_{\rm bg}$ is the sum of the appropriate radiative and non-radiative decay rates and $\Gamma^{\rm rot}_{\phi} = \Gamma_{\rm th} + \Gamma_{\rm st} + \Gamma_{\rm sc} + \Gamma_{\rm c}$ refers to the total dephasing rate for rotational superpositions (see Table~\ref{tab:parameters}).  Using the Maxwell-Bloch equations for an ensemble of uncoupled two-level systems interacting with a cavity \cite{Carmichael1994}, $\Gamma_{\rm{m}}$ in the limits of $C \ll1$ and $C\gg1$ is given by
\begin{equation}\label{eq:SNR}
\Gamma_{\rm{m}}=\frac{1}{n_{\rm{amp}}} \frac{M_0^2}{1+ C} \frac{g^2}{\kappa} \frac{\Gamma_1}{2\Gamma_2}.
\end{equation}
When the ensemble is optically 
thin ($C \ll 1$), the measurement rate is proportional to 
$M_0^2 \Gamma_\kappa$, the Purcell-enhanced radiation rate through the cavity; 
and to the partial width of the transition, $\Gamma_1/2\Gamma_2$.   When 
the ensemble is optically thick ($C \gg 1$), such as in the memory experiment, 
Eq.~\ref{eq:SNR} reduces to $\Gamma_{\rm{m}} \propto M_0\Gamma_1$, i.e. the maximum rate of 
photon scattering. Non-radiative decay (such as from collisions with buffer gas) does not reduce the detection rate but does increase the required measurement power.   Ideally, for detection the molecules would be on resonance with the cavity as this case is least sensitive to systematics; however, in principle the detection rate in Eq.~\ref{eq:SNR} can be attained even at large cavity-molecule detunings by using the appropriate measurement power.  

\section{Experimental scenarios and feasibility}
We formulate design goals for the three different classes of experiments, ranging from easiest (spectroscopy) to most difficult (qubit), with the memory experiment of intermediate complexity.  In general smaller trap widths are favorable for stronger coupling but are more technically challenging. Similarly, achieving the lowest temperatures is always favorable but experimentally more difficult.  The specific parameters envisioned for each class of experiment are given in Table~\ref{tab:parameters}.

Parameters for the spectroscopy experiment were chosen to be realizable with current technology, i.e. with moderately cold ($T_c\sim 1.5\,\rm{K}$) buffer gas for cooling and trap loading, a large ion trap, a modest $Q$ cavity, and as many molecules as possible ($N = n_{\rm i} \pi w^2 \lambda/2$).  Using helium buffer gas, all temperatures should be similar: $T_r \approx T_m \approx T_c$.  Even lower rotational temperatures would be advantageous, and may be possible via direct laser cooling \cite{Staanum10} and/or sympathetic cooling with co-located, laser cooled neutral atoms~\cite{Hudson09} or ions~\cite{vogelius_rotational_2006}.

\begin{table*}
\begin{tabular}{lcccc}
  \hline\hline
            &         &              & Experiment &  \\
  Parameter & [Units] & ~~~~Spectroscopy~~ &   Qubit   & Memory \\
    \hline\hline
  \multicolumn{5}{c}{Cavity, Trap, and Ensemble Parameters}\\
  \hline
  Trap Width*, $w$ & $[\rm{\mu m}]$ & $1000$  & $10$ & $10$ \\
  Trap Depth*, $\frac{U_{\rm{max}}}{e}$ & [mV] & 100 & 100 & 100 \\
  Drive Freq., $\frac{\Omega_{\rm{RF}}}{2\pi}$ & [MHz] & 1.3& 130& 130\\
  Trap Freq., $\frac{\omega_t}{2\pi}$ & [MHz] &  0.1 & 10 & 10 \\
  Configuration& &Large Cloud & Single Ion & Linear Chain \\
  Number of molecules*, $N$ & [1] & 170,000 & 1 & 1,700 \\
  Cryostat Temp.*, $T_c$ & [mK] & 1,500 & 20 & 20 \\
  Motional Temp.*, $T_{\rm{m}}$ & [mK] & 1,500 & $<1$ & $<1$ \\
  Rotational Temp.*, $T_{\rm{r}}$  & [mK] & 1,500 & $20$ & $20$ \\
  Cavity Q Factor*, $Q$ & [1] & $10^3$  &  $10^6$ & $10^6$ \\
  Cavity decay rate, $\frac{\kappa}{2\pi}$ & [Hz] & $9.0\times10^6$  & $9.0 \times 10^3$ & $9 \times 10^3$ \\
  \hline\hline
    \multicolumn{5}{c}{Decoherence Rates} \\
  \hline
  Suppression factor, $\Lambda$ & [1] & $1.5 \times 10^{-7}$ & $1.5\times10^{-3}$ & $1.5\times10^{-3}$ \\
  Thermal, $\frac{\Gamma_{\rm{th}}}{2\pi}$ & [Hz] & $4.6\times 10^{3}$ & 1 & 1 \\
  Stray field, $\frac{\Gamma_{\rm st}}{2 \pi}$ & [Hz] & $<1$ & $10$ & $10$ \\
  Space-charge, $\frac{\Gamma_{\rm{sc}}}{2\pi}$ & [Hz] & $4.6 \times 10^{4}$ & 0 & 0 \\
  Collisional, $\frac{\Gamma_{\rm{c}}}{2\pi}$ & [Hz] & $<1$ & 0 & $<1$\\
  Buffer gas*, $\frac{\Gamma_{\rm{bg}}}{2\pi}$ & [Hz] & $3.5 \times 10^{3}$ & 0 & 0\\
  Total rotational, $\frac{\Gamma^{\rm rot}_{\phi}}{2\pi}$ & [Hz]  & $51 \times 10^4$ & $10$ & $10$ \\
  HFS-encoded, $\frac{\Gamma^{\rm HFS}_{\phi}}{2\pi}$ & [Hz]  & & $0.1$ & $50$ \\
\hline\hline
      \multicolumn{5}{c}{Coupling strengths and detection rate} \\
  \hline
  Peak single-ion coupling $\frac{g_1}{2\pi}$ & [Hz] & $59$ & $5.9\times 10^3$ & $5.9\times 10^3$ \\
  Eff. single-ion coupling $\frac{g_1^{\rm eff}}{2\pi}$ & [Hz] & $30$ & $5.9\times 10^3$ & $30\times 10^3$ \\
  Eff. ensemble coupling $\frac{g^{\rm eff}_{N}}{2\pi} = \frac{g_1^{\rm eff}\sqrt{N}}{2\pi}$ & [Hz] & $6.5 \times 10^3$ &  & $170 \times 10^5$ \\
  Cooperativity, $C^{\rm eff}=\frac{2M_0(g^{\rm eff}_{N})^2}{\Gamma_{2} \kappa}$ & [1] & $1.8\times10^{-4}$ & $7.7\times10^2$& $6.4\times 10^5$ \\
  Cavity-enhanced reset rate\footnote{The decay rate can be controllably enhanced up to a maximum rate of $\kappa/2$.}, $\frac{\Gamma_{\kappa}}{2\pi}$ & [Hz] & $9.4$ & $7.8 \times 10^{3}$ & $4.5 \times 10^{3}$ \\
  Measurement Rate, $\frac{\Gamma_{\rm{m}}}{2 \pi}$ & [Hz] & $1.2 \times 10^{3}$ & $1.2 \times 10^{3}$ & $1.0  \times 10^{6}$ \\
\end{tabular}
\caption{Cavity, trap and ensemble parameters; coupling strengths; measurement rates; and decoherence rates for the three different experimental scenarios.  Parameters set by the experimental design are indicated by *; all other quantities are derived from these parameters and the common values $\beta = 0.5, q=0.2, \omega_{01} /2\pi= 4.5\,{\rm GHz}$, $m = 75$ a.m.u., $p = 3\,ea_0$, and $\Upsilon = 10^{-4}$. Effective couplings (denoted by superscript ``eff'') are adjusted for ensembles due to the fact that molecules may be located away from the midpoint along the length of the cavity, where the peak of the cavity electric field occurs; if atoms are distributed evenly throughout the length of the cavity, $g_1^{\rm eff} \approx g_1/2$.  All relaxation rates are for the first photon emitted by the molecules through the specified channel.  All quantities are reported to two significant digits, but true uncertainties are likely much larger. \label{tab:parameters}}
\end{table*}

To make an effective qubit or memory, the ion trap must be scaled down significantly and operated at dilution refrigerator temperatures ($T_c\sim 20\,\rm{mK}$) where there are no thermal photons in the cavity. In both these cases the rotational temperature can be brought into equilibrium with the environment (such that $T_r \approx T_c$) by coupling to the cavity, with equilibration rate $\Gamma_\kappa$.  Sufficiently low motional temperatures for the memory experiment ($T_m \lesssim T_c$) have been reached via sympathetic cooling with co-trapped laser cooled atomic ions \cite{molhave_formation_2000,Roth05}.  For the qubit experiment, the single molecular ion must be cooled to near its motional ground state (such that $T_m \ll T_c$) in which case Eq.~\ref{eq:thermaldephasing} for $\Gamma_{\rm th}$ is replaced by the motional heating rate, demonstrated to be as small as $\sim 1-100\,\rm{quanta/s}$~\cite{labaziewicz_suppression_2008} for similar cryogenic ion traps.   Sufficient cooling rates might be achievable via sympathetic cooling as used for atomic ions \cite{schmidt_spectroscopy_2005} or perhaps via cavity-assisted sideband cooling \cite{andre_coherent_2006} as envisioned for trapped neutral molecules.  The size of the memory is limited by the number of ions that can fit in the resonator along a linear chain, so that here $N \approx \lambda/(2n_i^{1/3})$.  If currently feasible superconducting cavity quality factors \cite{day_broadband_2003} can be maintained in the presence of the RF fields required for the ion trap \cite{seidelin_microfabricated_2006}, reaching the strong coupling regime appears achievable.    

Equipped with the measurement and decoherence rates, we can evaluate the feasibility of the three experimental scenarios.  The spectroscopy scenario is quite promising as a means for investigating rotational states of dipolar molecular ions.  Ensembles can have linewidths $\Gamma_2/2\pi \ll 1$~MHz, and can be detected at kHz rates (see  Table~\ref{tab:parameters}), likely enabling the measurement of rotational constants at the kHz level of accuracy.  The detection rate could be further improved by lowering the rotational temperature of the molecules, and/or using higher finesse cavities; spectral resolution could be improved by using smaller ensembles, by further cooling, and by reducing the buffer gas density.  Note that as described, the rotational transition frequency would have to be known fairly accurately in advance, in order to set the resonator frequency to be within a linewidth $\kappa$ of the molecular frequency.  However, it may be possible to add the ability to tune the resonator frequency e.g. by inserting a Josephson junction into the resonator~\cite{sandberg_tuningfield_2008}, or by using a mechanically tuned capacitor or quarter-wave shunt.  Such tunability would greatly enhance the flexibility of this spectroscopic method.

In evaluating the prospects for the quantum information experiments (qubit and quantum memory), a useful figure of merit is the number of coherent operations $N_{\rm{ops}}$ which can be performed. When the cavity resonant frequency is optimally detuned from the molecular frequency, this is given by $N_{\rm{ops}}=\sqrt{C}$ \cite{blais_quantum-information_2007}.  In computing $N_{\rm{ops}}$ we conservatively assume that HFS encoding is only used to increase the storage time rather than improving the gate fidelity.  With this assumption single molecule qubits could enable $ N_{\rm{ops}} \approx 30$ two-qubit gates, competitive with existing solid-state technologies (although with slower gate speed here).  In order for the memory to be viable, we have assumed in table~\ref{tab:parameters} that it has been cooled close to its motional ground state where its dephasing rate would then be limited by heating.  Cooling a large chain of ions to its ground state would require significant advance in cooling technology, which is also sought after for atomic ion quantum computation~\cite{lin_large-scale_2009}.  If it could be achieved the memory coupling strength would be comparable to solid-state qubit lifetimes;  hence the fidelity $\mathcal{F}$ with which information can be swapped from the ensemble to the cavity or to a co-located superconducting qubit \cite{rabl_hybrid_2006} is limited by $1-\mathcal{F} \approx \kappa/g_N \approx 5\%$.  This should be sufficient to enable a demonstration of a hybrid quantum memory.  In both the qubit and the memory experiments, the internal state of the system could be determined in much less than one lifetime ($\Gamma_2^{-1}$), making readout fidelity excellent.  If it were possible to maintain the HFS encoding protection while performing gates, fidelities could be improved further.  

\section{Conclusion}

These estimates show that it may be possible to realize an all-electrical interface to the rotational states of molecular ions.   This represents a promising approach for measuring the rotational structure of molecular ions, and could be interesting for demonstrating new methods applicable to quantum computing.   Using ions rather than neutral molecules allows easy trapping using known technology, while retaining reasonably large coupling strengths and keeping anticipated dephasing mechanisms under sufficient control.  As with neutral molecules, experiments attempting to realize qubits or quantum memories will be challenging; however initial demonstrations may be easier with ions than with neutrals.  Also as with neutrals, the use of molecular qubits opens some interesting new possibilities; for example, because molecules have both rotational and electronic transitions it may be possible to use this system as a single photon microwave-to-optical upconverter or downconverter.  Finally, our analysis may prove useful for evaluating likely linewidths and systematic errors in experiments that envision using precision spectroscopy of trapped dipolar molecular ions for tests of fundamental physics (see e.g. \cite{Flambaum07,Koelemeij07}). 

\begin{acknowledgments}
The authors would like to acknowledge useful discussions with P. Antohi, G. Akselrod, and J. Koch.  This work was supported in part by Yale University via the Quantum Information and Mesoscopic Physics Fellowship, and by the NSF Center for Ultracold Atoms, as well as grants DMR-0325580(NSF), DMR-0653377(NSF), and W911NF0510405(ARO).
\end{acknowledgments}


\begin{thebibliography}{57}%
\makeatletter
\providecommand \@ifxundefined [1]{%
 \@ifx{#1\undefined}
}%
\providecommand \@ifnum [1]{%
 \ifnum #1\expandafter \@firstoftwo
 \else \expandafter \@secondoftwo
 \fi
}%
\providecommand \@ifx [1]{%
 \ifx #1\expandafter \@firstoftwo
 \else \expandafter \@secondoftwo
 \fi
}%
\providecommand \natexlab [1]{#1}%
\providecommand \enquote  [1]{``#1''}%
\providecommand \bibnamefont  [1]{#1}%
\providecommand \bibfnamefont [1]{#1}%
\providecommand \citenamefont [1]{#1}%
\providecommand \href@noop [0]{\@secondoftwo}%
\providecommand \href [0]{\begingroup \@sanitize@url \@href}%
\providecommand \@href[1]{\@@startlink{#1}\@@href}%
\providecommand \@@href[1]{\endgroup#1\@@endlink}%
\providecommand \@sanitize@url [0]{\catcode `\\12\catcode `\$12\catcode
  `\&12\catcode `\#12\catcode `\^12\catcode `\_12\catcode `\%12\relax}%
\providecommand \@@startlink[1]{}%
\providecommand \@@endlink[0]{}%
\providecommand \url  [0]{\begingroup\@sanitize@url \@url }%
\providecommand \@url [1]{\endgroup\@href {#1}{\urlprefix }}%
\providecommand \urlprefix  [0]{URL }%
\providecommand \Eprint [0]{\href }%
\@ifxundefined \urlstyle {%
  \providecommand \doi  [0]{\begingroup \@sanitize@url \@doi}%
  \providecommand \@doi [1]{\endgroup \@@startlink {\doibase
  #1}doi:\discretionary {}{}{}#1\@@endlink }%
}{%
  \providecommand \doi  [0]{doi:\discretionary{}{}{}\begingroup
  \urlstyle{rm}\Url }%
}%
\providecommand \doibase [0]{http://dx.doi.org/}%
\providecommand \Doi [0]{\begingroup \@sanitize@url \@Doi }%
\providecommand \@Doi  [1]{\endgroup\@@startlink{\doibase#1}\@@Doi}%
\providecommand \@@Doi [1]{#1\@@endlink}%
\providecommand \selectlanguage [0]{\@gobble}%
\providecommand \bibinfo  [0]{\@secondoftwo}%
\providecommand \bibfield  [0]{\@secondoftwo}%
\providecommand \translation [1]{[#1]}%
\providecommand \BibitemOpen [0]{}%
\providecommand \bibitemStop [0]{}%
\providecommand \bibitemNoStop [0]{.\EOS\space}%
\providecommand \EOS [0]{\spacefactor3000\relax}%
\providecommand \BibitemShut  [1]{\csname bibitem#1\endcsname}%
\bibitem [{\citenamefont {Sauer}\ \emph {et~al.}(2005)\citenamefont {Sauer},
  \citenamefont {Hudson}, \citenamefont {Tarbutt},\ and\ \citenamefont
  {Hinds}}]{Sauer2005}%
  \BibitemOpen
  \bibfield  {author} {\bibinfo {author} {\bibfnamefont {B.}~\bibnamefont
  {Sauer}}, \bibinfo {author} {\bibfnamefont {J.}~\bibnamefont {Hudson}},
  \bibinfo {author} {\bibfnamefont {M.~R.}\ \bibnamefont {Tarbutt}}, \ and\
  \bibinfo {author} {\bibfnamefont {E.~A.}\ \bibnamefont {Hinds}},\ }\enquote
  {\bibinfo {title} {Cold molecules as a laboratory for particle physics},}\
  in\ \href@noop {} {\emph {\bibinfo {booktitle} {Interactions in Ultracold
  Gases}}},\ \bibinfo {editor} {edited by\ \bibinfo {editor} {\bibfnamefont
  {M.}~\bibnamefont {Weidemüller}}\ and\ \bibinfo {editor} {\bibfnamefont
  {C.}~\bibnamefont {Zimmermann}}}\ (\bibinfo  {publisher} {Wiley},\ \bibinfo
  {year} {2005})\ pp.\ \bibinfo {pages} {359--369}\BibitemShut {NoStop}%
\bibitem [{\citenamefont {Sinclair}\ \emph {et~al.}(2005)\citenamefont
  {Sinclair}, \citenamefont {Bohn}, \citenamefont {Leanhardt}, \citenamefont
  {Maletinsky}, \citenamefont {Meyer}, \citenamefont {Stutz},\ and\
  \citenamefont {Cornell}}]{Sinclair05}%
  \BibitemOpen
  \bibfield  {author} {\bibinfo {author} {\bibfnamefont {R.}~\bibnamefont
  {Sinclair}}, \bibinfo {author} {\bibfnamefont {J.}~\bibnamefont {Bohn}},
  \bibinfo {author} {\bibfnamefont {A.}~\bibnamefont {Leanhardt}}, \bibinfo
  {author} {\bibfnamefont {P.}~\bibnamefont {Maletinsky}}, \bibinfo {author}
  {\bibfnamefont {E.}~\bibnamefont {Meyer}}, \bibinfo {author} {\bibfnamefont
  {R.}~\bibnamefont {Stutz}}, \ and\ \bibinfo {author} {\bibfnamefont
  {E.}~\bibnamefont {Cornell}},\ }\href@noop {} {\bibfield  {journal} {\bibinfo
   {journal} {Bull. Am. Phys. Soc.},\ }\textbf {\bibinfo {volume} {50}},\
  \bibinfo {pages} {134} (\bibinfo {year} {2005})}\BibitemShut {NoStop}%
\bibitem [{\citenamefont {Schiller}\ and\ \citenamefont
  {Korobov}(2005)}]{Schiller05}%
  \BibitemOpen
  \bibfield  {author} {\bibinfo {author} {\bibfnamefont {S.}~\bibnamefont
  {Schiller}}\ and\ \bibinfo {author} {\bibfnamefont {V.}~\bibnamefont
  {Korobov}},\ }\href@noop {} {\bibfield  {journal} {\bibinfo  {journal} {Phys.
  Rev. A},\ }\textbf {\bibinfo {volume} {71}},\ \bibinfo {pages} {032505}
  (\bibinfo {year} {2005})}\BibitemShut {NoStop}%
\bibitem [{\citenamefont {DeMille}\ \emph
  {et~al.}(2008){\natexlab{a}}\citenamefont {DeMille}, \citenamefont {Sainis},
  \citenamefont {Sage}, \citenamefont {Bergeman}, \citenamefont {Kotochigova},\
  and\ \citenamefont {Tiesinga}}]{demille043202}%
  \BibitemOpen
  \bibfield  {author} {\bibinfo {author} {\bibfnamefont {D.}~\bibnamefont
  {DeMille}}, \bibinfo {author} {\bibfnamefont {S.}~\bibnamefont {Sainis}},
  \bibinfo {author} {\bibfnamefont {J.}~\bibnamefont {Sage}}, \bibinfo {author}
  {\bibfnamefont {T.}~\bibnamefont {Bergeman}}, \bibinfo {author}
  {\bibfnamefont {S.}~\bibnamefont {Kotochigova}}, \ and\ \bibinfo {author}
  {\bibfnamefont {E.}~\bibnamefont {Tiesinga}},\ }\href@noop {} {\bibfield
  {journal} {\bibinfo  {journal} {Phys. Rev. Lett.},\ }\textbf {\bibinfo
  {volume} {100}},\ \bibinfo {pages} {043202} (\bibinfo {year}
  {2008}{\natexlab{a}})}\BibitemShut {NoStop}%
\bibitem [{\citenamefont {Zelevinsky}\ \emph {et~al.}(2008)\citenamefont
  {Zelevinsky}, \citenamefont {Kotochigova},\ and\ \citenamefont
  {Ye}}]{zelevinsky043201}%
  \BibitemOpen
  \bibfield  {author} {\bibinfo {author} {\bibfnamefont {T.}~\bibnamefont
  {Zelevinsky}}, \bibinfo {author} {\bibfnamefont {S.}~\bibnamefont
  {Kotochigova}}, \ and\ \bibinfo {author} {\bibfnamefont {J.}~\bibnamefont
  {Ye}},\ }\href@noop {} {\bibfield  {journal} {\bibinfo  {journal} {Phys. Rev.
  Lett.},\ }\textbf {\bibinfo {volume} {100}},\ \bibinfo {pages} {043201}
  (\bibinfo {year} {2008})}\BibitemShut {NoStop}%
\bibitem [{\citenamefont {DeMille}\ \emph
  {et~al.}(2008){\natexlab{b}}\citenamefont {DeMille}, \citenamefont {Cahn},
  \citenamefont {Murphree}, \citenamefont {Rahmlow},\ and\ \citenamefont
  {Kozlov}}]{demille023003}%
  \BibitemOpen
  \bibfield  {author} {\bibinfo {author} {\bibfnamefont {D.}~\bibnamefont
  {DeMille}}, \bibinfo {author} {\bibfnamefont {S.~B.}\ \bibnamefont {Cahn}},
  \bibinfo {author} {\bibfnamefont {D.}~\bibnamefont {Murphree}}, \bibinfo
  {author} {\bibfnamefont {D.~A.}\ \bibnamefont {Rahmlow}}, \ and\ \bibinfo
  {author} {\bibfnamefont {M.~G.}\ \bibnamefont {Kozlov}},\ }\href@noop {}
  {\bibfield  {journal} {\bibinfo  {journal} {Phys. Rev. Lett.},\ }\textbf
  {\bibinfo {volume} {100}},\ \bibinfo {pages} {023003} (\bibinfo {year}
  {2008}{\natexlab{b}})}\BibitemShut {NoStop}%
\bibitem [{\citenamefont {Muller}\ \emph {et~al.}(2004)\citenamefont {Muller},
  \citenamefont {Herrmann}, \citenamefont {Saenz}, \citenamefont {Peters},\
  and\ \citenamefont {Lammerzahl}}]{muller_tests_2004}%
  \BibitemOpen
  \bibfield  {author} {\bibinfo {author} {\bibfnamefont {H.}~\bibnamefont
  {Muller}}, \bibinfo {author} {\bibfnamefont {S.}~\bibnamefont {Herrmann}},
  \bibinfo {author} {\bibfnamefont {A.}~\bibnamefont {Saenz}}, \bibinfo
  {author} {\bibfnamefont {A.}~\bibnamefont {Peters}}, \ and\ \bibinfo {author}
  {\bibfnamefont {C.}~\bibnamefont {Lammerzahl}},\ }\Doi
  {10.1103/PhysRevD.70.076004} {\bibfield  {journal} {\bibinfo  {journal}
  {Phys. Rev. D},\ }\textbf {\bibinfo {volume} {70}},\ \bibinfo {pages}
  {076004} (\bibinfo {year} {2004})}\BibitemShut {NoStop}%
\bibitem [{\citenamefont {DeMille}(2002)}]{demille_quantum_2002}%
  \BibitemOpen
  \bibfield  {author} {\bibinfo {author} {\bibfnamefont {D.}~\bibnamefont
  {DeMille}},\ }\href@noop {} {\bibfield  {journal} {\bibinfo  {journal} {Phys.
  Rev. Lett.},\ }\textbf {\bibinfo {volume} {88}},\ \bibinfo {pages} {067901}
  (\bibinfo {year} {2002})}\BibitemShut {NoStop}%
\bibitem [{\citenamefont {Andre}\ \emph {et~al.}(2006)\citenamefont {Andre},
  \citenamefont {DeMille}, \citenamefont {Doyle}, \citenamefont {Lukin},
  \citenamefont {Maxwell}, \citenamefont {Rabl}, \citenamefont {Schoelkopf},\
  and\ \citenamefont {Zoller}}]{andre_coherent_2006}%
  \BibitemOpen
  \bibfield  {author} {\bibinfo {author} {\bibfnamefont {A.}~\bibnamefont
  {Andre}}, \bibinfo {author} {\bibfnamefont {D.}~\bibnamefont {DeMille}},
  \bibinfo {author} {\bibfnamefont {J.~M.}\ \bibnamefont {Doyle}}, \bibinfo
  {author} {\bibfnamefont {M.~D.}\ \bibnamefont {Lukin}}, \bibinfo {author}
  {\bibfnamefont {S.~E.}\ \bibnamefont {Maxwell}}, \bibinfo {author}
  {\bibfnamefont {P.}~\bibnamefont {Rabl}}, \bibinfo {author} {\bibfnamefont
  {R.~J.}\ \bibnamefont {Schoelkopf}}, \ and\ \bibinfo {author} {\bibfnamefont
  {P.}~\bibnamefont {Zoller}},\ }\href@noop {} {\bibfield  {journal} {\bibinfo
  {journal} {Nature Phys.},\ }\textbf {\bibinfo {volume} {2}},\ \bibinfo
  {pages} {636} (\bibinfo {year} {2006})}\BibitemShut {NoStop}%
\bibitem [{\citenamefont {Rabl}\ \emph {et~al.}(2006)\citenamefont {Rabl},
  \citenamefont {DeMille}, \citenamefont {Doyle}, \citenamefont {Lukin},
  \citenamefont {Schoelkopf},\ and\ \citenamefont {Zoller}}]{rabl_hybrid_2006}%
  \BibitemOpen
  \bibfield  {author} {\bibinfo {author} {\bibfnamefont {P.}~\bibnamefont
  {Rabl}}, \bibinfo {author} {\bibfnamefont {D.}~\bibnamefont {DeMille}},
  \bibinfo {author} {\bibfnamefont {J.~M.}\ \bibnamefont {Doyle}}, \bibinfo
  {author} {\bibfnamefont {M.~D.}\ \bibnamefont {Lukin}}, \bibinfo {author}
  {\bibfnamefont {R.~J.}\ \bibnamefont {Schoelkopf}}, \ and\ \bibinfo {author}
  {\bibfnamefont {P.}~\bibnamefont {Zoller}},\ }\href@noop {} {\bibfield
  {journal} {\bibinfo  {journal} {Phys. Rev. Lett.},\ }\textbf {\bibinfo
  {volume} {97}},\ \bibinfo {pages} {033003} (\bibinfo {year}
  {2006})}\BibitemShut {NoStop}%
\bibitem [{\citenamefont {Balakrishnan}\ and\ \citenamefont
  {Dalgarno}(2001)}]{balakrishnan_chemistry_2001}%
  \BibitemOpen
  \bibfield  {author} {\bibinfo {author} {\bibfnamefont {N.}~\bibnamefont
  {Balakrishnan}}\ and\ \bibinfo {author} {\bibfnamefont {A.}~\bibnamefont
  {Dalgarno}},\ }\href@noop {} {\bibfield  {journal} {\bibinfo  {journal}
  {Chem. Phys. Lett.},\ }\textbf {\bibinfo {volume} {341}},\ \bibinfo {pages}
  {652} (\bibinfo {year} {2001})}\BibitemShut {NoStop}%
\bibitem [{\citenamefont {Shuman}\ \emph {et~al.}(2009)\citenamefont {Shuman},
  \citenamefont {Barry}, \citenamefont {Glenn},\ and\ \citenamefont
  {DeMille}}]{Shuman09}%
  \BibitemOpen
  \bibfield  {author} {\bibinfo {author} {\bibfnamefont {E.~S.}\ \bibnamefont
  {Shuman}}, \bibinfo {author} {\bibfnamefont {J.~F.}\ \bibnamefont {Barry}},
  \bibinfo {author} {\bibfnamefont {D.~R.}\ \bibnamefont {Glenn}}, \ and\
  \bibinfo {author} {\bibfnamefont {D.}~\bibnamefont {DeMille}},\ }\href@noop
  {} {\bibfield  {journal} {\bibinfo  {journal} {Phys. Rev. Lett.},\ }\textbf
  {\bibinfo {volume} {103}},\ \bibinfo {pages} {223001} (\bibinfo {year}
  {2009})}\BibitemShut {NoStop}%
\bibitem [{\citenamefont {Ni}\ \emph {et~al.}(2008)\citenamefont {Ni},
  \citenamefont {Ospelkaus}, \citenamefont {de~Miranda}, \citenamefont {Pe'er},
  \citenamefont {Neyenhuis}, \citenamefont {Zirbel}, \citenamefont
  {Kotochigova}, \citenamefont {Julienne}, \citenamefont {Jin},\ and\
  \citenamefont {Ye}}]{Ni09182008}%
  \BibitemOpen
  \bibfield  {author} {\bibinfo {author} {\bibfnamefont {K.-K.}\ \bibnamefont
  {Ni}}, \bibinfo {author} {\bibfnamefont {S.}~\bibnamefont {Ospelkaus}},
  \bibinfo {author} {\bibfnamefont {M.~H.~G.}\ \bibnamefont {de~Miranda}},
  \bibinfo {author} {\bibfnamefont {A.}~\bibnamefont {Pe'er}}, \bibinfo
  {author} {\bibfnamefont {B.}~\bibnamefont {Neyenhuis}}, \bibinfo {author}
  {\bibfnamefont {J.~J.}\ \bibnamefont {Zirbel}}, \bibinfo {author}
  {\bibfnamefont {S.}~\bibnamefont {Kotochigova}}, \bibinfo {author}
  {\bibfnamefont {P.~S.}\ \bibnamefont {Julienne}}, \bibinfo {author}
  {\bibfnamefont {D.~S.}\ \bibnamefont {Jin}}, \ and\ \bibinfo {author}
  {\bibfnamefont {J.}~\bibnamefont {Ye}},\ }\href@noop {} {\bibfield  {journal}
  {\bibinfo  {journal} {Science},\ }\textbf {\bibinfo {volume} {322}},\
  \bibinfo {pages} {231} (\bibinfo {year} {2008})}\BibitemShut {NoStop}%
\bibitem [{\citenamefont {Bethlem}\ \emph {et~al.}(2000)\citenamefont
  {Bethlem}, \citenamefont {Berden}, \citenamefont {Crompvoets}, \citenamefont
  {Jongma}, \citenamefont {van Roij},\ and\ \citenamefont
  {Meijer}}]{bethlem_electrostatic_2000}%
  \BibitemOpen
  \bibfield  {author} {\bibinfo {author} {\bibfnamefont {H.~L.}\ \bibnamefont
  {Bethlem}}, \bibinfo {author} {\bibfnamefont {G.}~\bibnamefont {Berden}},
  \bibinfo {author} {\bibfnamefont {F.~M.~H.}\ \bibnamefont {Crompvoets}},
  \bibinfo {author} {\bibfnamefont {R.~T.}\ \bibnamefont {Jongma}}, \bibinfo
  {author} {\bibfnamefont {A.~J.~A.}\ \bibnamefont {van Roij}}, \ and\ \bibinfo
  {author} {\bibfnamefont {G.}~\bibnamefont {Meijer}},\ }\href@noop {}
  {\bibfield  {journal} {\bibinfo  {journal} {Nature},\ }\textbf {\bibinfo
  {volume} {406}},\ \bibinfo {pages} {491} (\bibinfo {year}
  {2000})}\BibitemShut {NoStop}%
\bibitem [{\citenamefont {Weinstein}\ \emph {et~al.}(1998)\citenamefont
  {Weinstein}, \citenamefont {deCarvalho}, \citenamefont {Guillet},
  \citenamefont {Friedrich},\ and\ \citenamefont
  {Doyle}}]{weinstein_magnetic_1998}%
  \BibitemOpen
  \bibfield  {author} {\bibinfo {author} {\bibfnamefont {J.~D.}\ \bibnamefont
  {Weinstein}}, \bibinfo {author} {\bibfnamefont {R.}~\bibnamefont
  {deCarvalho}}, \bibinfo {author} {\bibfnamefont {T.}~\bibnamefont {Guillet}},
  \bibinfo {author} {\bibfnamefont {B.}~\bibnamefont {Friedrich}}, \ and\
  \bibinfo {author} {\bibfnamefont {J.~M.}\ \bibnamefont {Doyle}},\ }\href@noop
  {} {\bibfield  {journal} {\bibinfo  {journal} {Nature},\ }\textbf {\bibinfo
  {volume} {395}},\ \bibinfo {pages} {148} (\bibinfo {year}
  {1998})}\BibitemShut {NoStop}%
\bibitem [{\citenamefont {Molhave}\ and\ \citenamefont
  {Drewsen}(2000)}]{molhave_formation_2000}%
  \BibitemOpen
  \bibfield  {author} {\bibinfo {author} {\bibfnamefont {K.}~\bibnamefont
  {Molhave}}\ and\ \bibinfo {author} {\bibfnamefont {M.}~\bibnamefont
  {Drewsen}},\ }\href@noop {} {\bibfield  {journal} {\bibinfo  {journal} {Phys.
  Rev. A},\ }\textbf {\bibinfo {volume} {62}},\ \bibinfo {pages} {011401}
  (\bibinfo {year} {2000})}\BibitemShut {NoStop}%
\bibitem [{\citenamefont {Roth}\ \emph {et~al.}(2005)\citenamefont {Roth},
  \citenamefont {Ostendorf}, \citenamefont {Wenz},\ and\ \citenamefont
  {Schiller}}]{Roth05}%
  \BibitemOpen
  \bibfield  {author} {\bibinfo {author} {\bibfnamefont {B.}~\bibnamefont
  {Roth}}, \bibinfo {author} {\bibfnamefont {A.}~\bibnamefont {Ostendorf}},
  \bibinfo {author} {\bibfnamefont {H.}~\bibnamefont {Wenz}}, \ and\ \bibinfo
  {author} {\bibfnamefont {S.}~\bibnamefont {Schiller}},\ }\href@noop {}
  {\bibfield  {journal} {\bibinfo  {journal} {J. Phys. B},\ }\textbf {\bibinfo
  {volume} {38}},\ \bibinfo {pages} {3673} (\bibinfo {year}
  {2005})}\BibitemShut {NoStop}%
\bibitem [{\citenamefont {Krems}\ \emph {et~al.}(2009)\citenamefont {Krems},
  \citenamefont {Stwalley},\ and\ \citenamefont
  {Friedrich}}]{KremsColdMoleculesReview}%
  \BibitemOpen
  \bibinfo {editor} {\bibfnamefont {R.~V.}\ \bibnamefont {Krems}}, \bibinfo
  {editor} {\bibfnamefont {W.~C.}\ \bibnamefont {Stwalley}}, \ and\ \bibinfo
  {editor} {\bibfnamefont {B.}~\bibnamefont {Friedrich}},\ eds.,\ \href@noop {}
  {\emph {\bibinfo {title} {Cold Molecules: Theory, Experiment,
  Applications}}}\ (\bibinfo  {publisher} {CRC Press},\ \bibinfo {address}
  {Boca Raton},\ \bibinfo {year} {2009})\BibitemShut {NoStop}%
\bibitem [{\citenamefont {Carr}\ \emph {et~al.}(2009)\citenamefont {Carr},
  \citenamefont {DeMille}, \citenamefont {Krems},\ and\ \citenamefont
  {Ye}}]{Carr09}%
  \BibitemOpen
  \bibfield  {author} {\bibinfo {author} {\bibfnamefont {L.~D.}\ \bibnamefont
  {Carr}}, \bibinfo {author} {\bibfnamefont {D.}~\bibnamefont {DeMille}},
  \bibinfo {author} {\bibfnamefont {R.~V.}\ \bibnamefont {Krems}}, \ and\
  \bibinfo {author} {\bibfnamefont {J.}~\bibnamefont {Ye}},\ }\href@noop {}
  {\bibfield  {journal} {\bibinfo  {journal} {New J. Phys.},\ }\textbf
  {\bibinfo {volume} {11}},\ \bibinfo {pages} {055049} (\bibinfo {year}
  {2009})}\BibitemShut {NoStop}%
\bibitem [{\citenamefont {Tian}\ \emph {et~al.}(2004)\citenamefont {Tian},
  \citenamefont {Rabl}, \citenamefont {Blatt},\ and\ \citenamefont
  {Zoller}}]{tian_interfacing_2004}%
  \BibitemOpen
  \bibfield  {author} {\bibinfo {author} {\bibfnamefont {L.}~\bibnamefont
  {Tian}}, \bibinfo {author} {\bibfnamefont {P.}~\bibnamefont {Rabl}}, \bibinfo
  {author} {\bibfnamefont {R.}~\bibnamefont {Blatt}}, \ and\ \bibinfo {author}
  {\bibfnamefont {P.}~\bibnamefont {Zoller}},\ }\href@noop {} {\bibfield
  {journal} {\bibinfo  {journal} {Phys. Rev. Lett.},\ }\textbf {\bibinfo
  {volume} {92}},\ \bibinfo {pages} {247902} (\bibinfo {year}
  {2004})}\BibitemShut {NoStop}%
\bibitem [{\citenamefont {Tian}\ and\ \citenamefont
  {Zoller}(2004)}]{tian_coupled_2004}%
  \BibitemOpen
  \bibfield  {author} {\bibinfo {author} {\bibfnamefont {L.}~\bibnamefont
  {Tian}}\ and\ \bibinfo {author} {\bibfnamefont {P.}~\bibnamefont {Zoller}},\
  }\href@noop {} {\bibfield  {journal} {\bibinfo  {journal} {Phys. Rev.
  Lett.},\ }\textbf {\bibinfo {volume} {93}},\ \bibinfo {pages} {266403}
  (\bibinfo {year} {2004})}\BibitemShut {NoStop}%
\bibitem [{\citenamefont {Verdu}\ \emph {et~al.}(2009)\citenamefont {Verdu},
  \citenamefont {Zoubi}, \citenamefont {Koller}, \citenamefont {Majer},
  \citenamefont {Ritsch},\ and\ \citenamefont
  {Schmiedmayer}}]{verdu_strong_2009}%
  \BibitemOpen
  \bibfield  {author} {\bibinfo {author} {\bibfnamefont {J.}~\bibnamefont
  {Verdu}}, \bibinfo {author} {\bibfnamefont {H.}~\bibnamefont {Zoubi}},
  \bibinfo {author} {\bibfnamefont {{\relax Ch}.}~\bibnamefont {Koller}},
  \bibinfo {author} {\bibfnamefont {J.}~\bibnamefont {Majer}}, \bibinfo
  {author} {\bibfnamefont {H.}~\bibnamefont {Ritsch}}, \ and\ \bibinfo {author}
  {\bibfnamefont {J.}~\bibnamefont {Schmiedmayer}},\ }\href@noop {} {\bibfield
  {journal} {\bibinfo  {journal} {Phys. Rev. Lett.},\ }\textbf {\bibinfo
  {volume} {103}},\ \bibinfo {pages} {043603} (\bibinfo {year}
  {2009})}\BibitemShut {NoStop}%
\bibitem [{\citenamefont {Imamo\u{g}lu}(2009)}]{imamoglu_cavity_2009}%
  \BibitemOpen
  \bibfield  {author} {\bibinfo {author} {\bibfnamefont {A.}~\bibnamefont
  {Imamo\u{g}lu}},\ }\href@noop {} {\bibfield  {journal} {\bibinfo  {journal}
  {Phys. Rev. Lett.},\ }\textbf {\bibinfo {volume} {102}},\ \bibinfo {pages}
  {083602} (\bibinfo {year} {2009})}\BibitemShut {NoStop}%
\bibitem [{\citenamefont {Rabl}\ and\ \citenamefont
  {Zoller}(2007)}]{rabl_molecular_2007}%
  \BibitemOpen
  \bibfield  {author} {\bibinfo {author} {\bibfnamefont {P.}~\bibnamefont
  {Rabl}}\ and\ \bibinfo {author} {\bibfnamefont {P.}~\bibnamefont {Zoller}},\
  }\href@noop {} {\bibfield  {journal} {\bibinfo  {journal} {Phys. Rev. A},\
  }\textbf {\bibinfo {volume} {76}},\ \bibinfo {pages} {042308} (\bibinfo
  {year} {2007})}\BibitemShut {NoStop}%
\bibitem [{\citenamefont {Meek}\ \emph {et~al.}(2009)\citenamefont {Meek},
  \citenamefont {Conrad},\ and\ \citenamefont {Meijer}}]{Meek2009}%
  \BibitemOpen
  \bibfield  {author} {\bibinfo {author} {\bibfnamefont {S.~A.}\ \bibnamefont
  {Meek}}, \bibinfo {author} {\bibfnamefont {H.}~\bibnamefont {Conrad}}, \ and\
  \bibinfo {author} {\bibfnamefont {G.}~\bibnamefont {Meijer}},\ }\href@noop {}
  {\bibfield  {journal} {\bibinfo  {journal} {Science},\ }\textbf {\bibinfo
  {volume} {324}},\ \bibinfo {pages} {1699} (\bibinfo {year}
  {2009})}\BibitemShut {NoStop}%
\bibitem [{\citenamefont {Chiaverini}\ \emph {et~al.}(2005)\citenamefont
  {Chiaverini}, \citenamefont {Blakestad}, \citenamefont {Britton},
  \citenamefont {Jost}, \citenamefont {Langer}, \citenamefont {Leibfried},
  \citenamefont {Ozeri},\ and\ \citenamefont
  {Wineland}}]{chiaverini_surface-electrode_2005}%
  \BibitemOpen
  \bibfield  {author} {\bibinfo {author} {\bibfnamefont {J.}~\bibnamefont
  {Chiaverini}}, \bibinfo {author} {\bibfnamefont {R.~B.}\ \bibnamefont
  {Blakestad}}, \bibinfo {author} {\bibfnamefont {J.}~\bibnamefont {Britton}},
  \bibinfo {author} {\bibfnamefont {J.~D.}\ \bibnamefont {Jost}}, \bibinfo
  {author} {\bibfnamefont {C.}~\bibnamefont {Langer}}, \bibinfo {author}
  {\bibfnamefont {D.}~\bibnamefont {Leibfried}}, \bibinfo {author}
  {\bibfnamefont {R.}~\bibnamefont {Ozeri}}, \ and\ \bibinfo {author}
  {\bibfnamefont {D.~J.}\ \bibnamefont {Wineland}},\ }\href@noop {} {\bibfield
  {journal} {\bibinfo  {journal} {Quantum Information \& Computation},\
  }\textbf {\bibinfo {volume} {5}},\ \bibinfo {pages} {419} (\bibinfo {year}
  {2005})}\BibitemShut {NoStop}%
\bibitem [{\citenamefont {Seidelin}\ \emph {et~al.}(2006)\citenamefont
  {Seidelin}, \citenamefont {Chiaverini}, \citenamefont {Reichle},
  \citenamefont {Bollinger}, \citenamefont {Leibfried}, \citenamefont
  {Britton}, \citenamefont {Wesenberg}, \citenamefont {Blakestad},
  \citenamefont {Epstein}, \citenamefont {Hume}, \citenamefont {Itano},
  \citenamefont {Jost}, \citenamefont {Langer}, \citenamefont {Ozeri},
  \citenamefont {Shiga},\ and\ \citenamefont
  {Wineland}}]{seidelin_microfabricated_2006}%
  \BibitemOpen
  \bibfield  {author} {\bibinfo {author} {\bibfnamefont {S.}~\bibnamefont
  {Seidelin}}, \bibinfo {author} {\bibfnamefont {J.}~\bibnamefont
  {Chiaverini}}, \bibinfo {author} {\bibfnamefont {R.}~\bibnamefont {Reichle}},
  \bibinfo {author} {\bibfnamefont {J.~J.}\ \bibnamefont {Bollinger}}, \bibinfo
  {author} {\bibfnamefont {D.}~\bibnamefont {Leibfried}}, \bibinfo {author}
  {\bibfnamefont {J.}~\bibnamefont {Britton}}, \bibinfo {author} {\bibfnamefont
  {J.~H.}\ \bibnamefont {Wesenberg}}, \bibinfo {author} {\bibfnamefont {R.~B.}\
  \bibnamefont {Blakestad}}, \bibinfo {author} {\bibfnamefont {R.~J.}\
  \bibnamefont {Epstein}}, \bibinfo {author} {\bibfnamefont {D.~B.}\
  \bibnamefont {Hume}}, \bibinfo {author} {\bibfnamefont {W.~M.}\ \bibnamefont
  {Itano}}, \bibinfo {author} {\bibfnamefont {J.~D.}\ \bibnamefont {Jost}},
  \bibinfo {author} {\bibfnamefont {C.}~\bibnamefont {Langer}}, \bibinfo
  {author} {\bibfnamefont {R.}~\bibnamefont {Ozeri}}, \bibinfo {author}
  {\bibfnamefont {N.}~\bibnamefont {Shiga}}, \ and\ \bibinfo {author}
  {\bibfnamefont {D.~J.}\ \bibnamefont {Wineland}},\ }\href@noop {} {\bibfield
  {journal} {\bibinfo  {journal} {Phys. Rev. Lett.},\ }\textbf {\bibinfo
  {volume} {96}},\ \bibinfo {pages} {253003} (\bibinfo {year}
  {2006})}\BibitemShut {NoStop}%
\bibitem [{\citenamefont {Drewsen}\ \emph {et~al.}(1998)\citenamefont
  {Drewsen}, \citenamefont {Brodersen}, \citenamefont {Hornekaer},
  \citenamefont {Hangst},\ and\ \citenamefont {Schiffer}}]{drewsen_large_1998}%
  \BibitemOpen
  \bibfield  {author} {\bibinfo {author} {\bibfnamefont {M.}~\bibnamefont
  {Drewsen}}, \bibinfo {author} {\bibfnamefont {C.}~\bibnamefont {Brodersen}},
  \bibinfo {author} {\bibfnamefont {L.}~\bibnamefont {Hornekaer}}, \bibinfo
  {author} {\bibfnamefont {J.~S.}\ \bibnamefont {Hangst}}, \ and\ \bibinfo
  {author} {\bibfnamefont {J.~P.}\ \bibnamefont {Schiffer}},\ }\href@noop {}
  {\bibfield  {journal} {\bibinfo  {journal} {Phys. Rev. Lett.},\ }\textbf
  {\bibinfo {volume} {81}},\ \bibinfo {pages} {2878} (\bibinfo {year}
  {1998})}\BibitemShut {NoStop}%
\bibitem [{\citenamefont {Vion}\ \emph {et~al.}(2002)\citenamefont {Vion},
  \citenamefont {Aassime}, \citenamefont {Cottet}, \citenamefont {Joyez},
  \citenamefont {Pothier}, \citenamefont {Urbina}, \citenamefont {Esteve},\
  and\ \citenamefont {Devoret}}]{vion_manipulatingquantum_2002}%
  \BibitemOpen
  \bibfield  {author} {\bibinfo {author} {\bibfnamefont {D.}~\bibnamefont
  {Vion}}, \bibinfo {author} {\bibfnamefont {A.}~\bibnamefont {Aassime}},
  \bibinfo {author} {\bibfnamefont {A.}~\bibnamefont {Cottet}}, \bibinfo
  {author} {\bibfnamefont {P.}~\bibnamefont {Joyez}}, \bibinfo {author}
  {\bibfnamefont {H.}~\bibnamefont {Pothier}}, \bibinfo {author} {\bibfnamefont
  {C.}~\bibnamefont {Urbina}}, \bibinfo {author} {\bibfnamefont
  {D.}~\bibnamefont {Esteve}}, \ and\ \bibinfo {author} {\bibfnamefont {M.~H.}\
  \bibnamefont {Devoret}},\ }\href@noop {} {\bibfield  {journal} {\bibinfo
  {journal} {Science},\ }\textbf {\bibinfo {volume} {296}},\ \bibinfo {pages}
  {886} (\bibinfo {year} {2002})}\BibitemShut {NoStop}%
\bibitem [{\citenamefont {Wing}(1988)}]{Wing1988}%
  \BibitemOpen
  \bibfield  {author} {\bibinfo {author} {\bibfnamefont {W.~H.}\ \bibnamefont
  {Wing}},\ }\href@noop {} {\bibfield  {journal} {\bibinfo  {journal} {Philos.
  Trans. R. Soc. Lond. A},\ }\textbf {\bibinfo {volume} {324}},\ \bibinfo
  {pages} {75} (\bibinfo {year} {1988})}\BibitemShut {NoStop}%
\bibitem [{\citenamefont {Asvany}\ \emph {et~al.}(2008)\citenamefont {Asvany},
  \citenamefont {Ricken}, \citenamefont {M\"uller}, \citenamefont {Wiedner},
  \citenamefont {Giesen},\ and\ \citenamefont {Schlemmer}}]{Asvany08}%
  \BibitemOpen
  \bibfield  {author} {\bibinfo {author} {\bibfnamefont {O.}~\bibnamefont
  {Asvany}}, \bibinfo {author} {\bibfnamefont {O.}~\bibnamefont {Ricken}},
  \bibinfo {author} {\bibfnamefont {H.~S.~P.}\ \bibnamefont {M\"uller}},
  \bibinfo {author} {\bibfnamefont {M.~C.}\ \bibnamefont {Wiedner}}, \bibinfo
  {author} {\bibfnamefont {T.~F.}\ \bibnamefont {Giesen}}, \ and\ \bibinfo
  {author} {\bibfnamefont {S.}~\bibnamefont {Schlemmer}},\ }\href@noop {}
  {\bibfield  {journal} {\bibinfo  {journal} {Phys. Rev. Lett.},\ }\textbf
  {\bibinfo {volume} {100}},\ \bibinfo {pages} {233004} (\bibinfo {year}
  {2008})}\BibitemShut {NoStop}%
\bibitem [{\citenamefont {Koelemeij}\ \emph {et~al.}(2007)\citenamefont
  {Koelemeij}, \citenamefont {Roth}, \citenamefont {Wicht}, \citenamefont
  {Ernsting},\ and\ \citenamefont {Schiller}}]{Koelemeij07}%
  \BibitemOpen
  \bibfield  {author} {\bibinfo {author} {\bibfnamefont {J.~C.~J.}\
  \bibnamefont {Koelemeij}}, \bibinfo {author} {\bibfnamefont {B.}~\bibnamefont
  {Roth}}, \bibinfo {author} {\bibfnamefont {A.}~\bibnamefont {Wicht}},
  \bibinfo {author} {\bibfnamefont {I.}~\bibnamefont {Ernsting}}, \ and\
  \bibinfo {author} {\bibfnamefont {S.}~\bibnamefont {Schiller}},\ }\href@noop
  {} {\bibfield  {journal} {\bibinfo  {journal} {Phys. Rev. Lett.},\ }\textbf
  {\bibinfo {volume} {98}},\ \bibinfo {eid} {173002} (\bibinfo {year}
  {2007})}\BibitemShut {NoStop}%
\bibitem [{\citenamefont {Huber}\ and\ \citenamefont
  {Herzberg}(1979)}]{Herzbergconstants}%
  \BibitemOpen
  \bibfield  {author} {\bibinfo {author} {\bibfnamefont {K.~P.}\ \bibnamefont
  {Huber}}\ and\ \bibinfo {author} {\bibfnamefont {G.}~\bibnamefont
  {Herzberg}},\ }\href@noop {} {\emph {\bibinfo {title} {Molecular Spectra and
  Molecular Structure IV: Constants of Diatomic Molecules}}}\ (\bibinfo
  {publisher} {Van Nostrand Reinhold},\ \bibinfo {address} {New York},\
  \bibinfo {year} {1979})\BibitemShut {NoStop}%
\bibitem [{\citenamefont {Saykally}\ and\ \citenamefont
  {Woods}(1981)}]{Saykally81}%
  \BibitemOpen
  \bibfield  {author} {\bibinfo {author} {\bibfnamefont {R.~J.}\ \bibnamefont
  {Saykally}}\ and\ \bibinfo {author} {\bibfnamefont {R.~C.}\ \bibnamefont
  {Woods}},\ }\href@noop {} {\bibfield  {journal} {\bibinfo  {journal} {Ann.
  Rev. Phys. Chem.},\ }\textbf {\bibinfo {volume} {32}},\ \bibinfo {pages}
  {403} (\bibinfo {year} {1981})}\BibitemShut {NoStop}%
\bibitem [{\citenamefont {Hirota}(1992)}]{Hirota92}%
  \BibitemOpen
  \bibfield  {author} {\bibinfo {author} {\bibfnamefont {E.}~\bibnamefont
  {Hirota}},\ }\href@noop {} {\bibfield  {journal} {\bibinfo  {journal} {Chem.
  Rev.},\ }\textbf {\bibinfo {volume} {92}},\ \bibinfo {pages} {141} (\bibinfo
  {year} {1992})}\BibitemShut {NoStop}%
\bibitem [{\citenamefont {Walls}\ and\ \citenamefont
  {Milburn}(2008)}]{WallsMilburn}%
  \BibitemOpen
  \bibfield  {author} {\bibinfo {author} {\bibfnamefont {D.}~\bibnamefont
  {Walls}}\ and\ \bibinfo {author} {\bibfnamefont {G.~J.}\ \bibnamefont
  {Milburn}},\ }\href@noop {} {\emph {\bibinfo {title} {Quantum Optics}}}\
  (\bibinfo  {publisher} {Springer},\ \bibinfo {year} {2008})\ Chap.\ \bibinfo
  {chapter} {CQED}, pp.\ \bibinfo {pages} {213--227}\BibitemShut {NoStop}%
\bibitem [{\citenamefont {Day}\ \emph {et~al.}(2003)\citenamefont {Day},
  \citenamefont {LeDuc}, \citenamefont {Mazin}, \citenamefont {Vayonakis},\
  and\ \citenamefont {Zmuidzinas}}]{day_broadband_2003}%
  \BibitemOpen
  \bibfield  {author} {\bibinfo {author} {\bibfnamefont {P.~K.}\ \bibnamefont
  {Day}}, \bibinfo {author} {\bibfnamefont {H.~G.}\ \bibnamefont {LeDuc}},
  \bibinfo {author} {\bibfnamefont {B.~A.}\ \bibnamefont {Mazin}}, \bibinfo
  {author} {\bibfnamefont {A.}~\bibnamefont {Vayonakis}}, \ and\ \bibinfo
  {author} {\bibfnamefont {J.}~\bibnamefont {Zmuidzinas}},\ }\href@noop {}
  {\bibfield  {journal} {\bibinfo  {journal} {Nature},\ }\textbf {\bibinfo
  {volume} {425}},\ \bibinfo {pages} {817} (\bibinfo {year}
  {2003})}\BibitemShut {NoStop}%
\bibitem [{\citenamefont {Raouafi}\ \emph {et~al.}(1999)\citenamefont
  {Raouafi}, \citenamefont {Jeung},\ and\ \citenamefont
  {Jungen}}]{Raouafi1999248}%
  \BibitemOpen
  \bibfield  {author} {\bibinfo {author} {\bibfnamefont {S.}~\bibnamefont
  {Raouafi}}, \bibinfo {author} {\bibfnamefont {G.-H.}\ \bibnamefont {Jeung}},
  \ and\ \bibinfo {author} {\bibfnamefont {{\relax Ch}.}~\bibnamefont
  {Jungen}},\ }\href@noop {} {\bibfield  {journal} {\bibinfo  {journal} {J.
  Mol. Spectrosc.},\ }\textbf {\bibinfo {volume} {196}},\ \bibinfo {pages}
  {248} (\bibinfo {year} {1999})}\BibitemShut {NoStop}%
\bibitem [{\citenamefont {Kotochigova}(2008)}]{Kotochigova2008}%
  \BibitemOpen
  \bibfield  {author} {\bibinfo {author} {\bibfnamefont {S.}~\bibnamefont
  {Kotochigova}},\ }\href@noop {} {}\bibinfo {howpublished} {private
  communication} (\bibinfo {year} {2008})\BibitemShut {NoStop}%
\bibitem [{\citenamefont {Schoelkopf}\ and\ \citenamefont
  {Girvin}(2008)}]{schoelkopf_wiring_2008}%
  \BibitemOpen
  \bibfield  {author} {\bibinfo {author} {\bibfnamefont {R.~J.}\ \bibnamefont
  {Schoelkopf}}\ and\ \bibinfo {author} {\bibfnamefont {S.~M.}\ \bibnamefont
  {Girvin}},\ }\href@noop {} {\bibfield  {journal} {\bibinfo  {journal}
  {Nature},\ }\textbf {\bibinfo {volume} {451}},\ \bibinfo {pages} {664}
  (\bibinfo {year} {2008})}\BibitemShut {NoStop}%
\bibitem [{\citenamefont {Tavis}\ and\ \citenamefont
  {Cummings}(1968)}]{tavis_exact_1968}%
  \BibitemOpen
  \bibfield  {author} {\bibinfo {author} {\bibfnamefont {M.}~\bibnamefont
  {Tavis}}\ and\ \bibinfo {author} {\bibfnamefont {F.~W.}\ \bibnamefont
  {Cummings}},\ }\href@noop {} {\bibfield  {journal} {\bibinfo  {journal}
  {Phys. Rev.},\ }\textbf {\bibinfo {volume} {170}},\ \bibinfo {pages} {379}
  (\bibinfo {year} {1968})}\BibitemShut {NoStop}%
\bibitem [{\citenamefont {Sandberg}\ \emph {et~al.}(2008)\citenamefont
  {Sandberg}, \citenamefont {Wilson}, \citenamefont {Persson}, \citenamefont
  {Bauch}, \citenamefont {Johansson}, \citenamefont {Shumeiko}, \citenamefont
  {Duty},\ and\ \citenamefont {Delsing}}]{sandberg_tuningfield_2008}%
  \BibitemOpen
  \bibfield  {author} {\bibinfo {author} {\bibfnamefont {M.}~\bibnamefont
  {Sandberg}}, \bibinfo {author} {\bibfnamefont {C.~M.}\ \bibnamefont
  {Wilson}}, \bibinfo {author} {\bibfnamefont {F.}~\bibnamefont {Persson}},
  \bibinfo {author} {\bibfnamefont {T.}~\bibnamefont {Bauch}}, \bibinfo
  {author} {\bibfnamefont {G.}~\bibnamefont {Johansson}}, \bibinfo {author}
  {\bibfnamefont {V.}~\bibnamefont {Shumeiko}}, \bibinfo {author}
  {\bibfnamefont {T.}~\bibnamefont {Duty}}, \ and\ \bibinfo {author}
  {\bibfnamefont {P.}~\bibnamefont {Delsing}},\ }\href@noop {} {\bibfield
  {journal} {\bibinfo  {journal} {Appl. Phys. Lett.},\ }\textbf {\bibinfo
  {volume} {92}},\ \bibinfo {pages} {203501} (\bibinfo {year}
  {2008})}\BibitemShut {NoStop}%
\bibitem [{\citenamefont {Paul}(1990)}]{paul_electromagnetic_1990}%
  \BibitemOpen
  \bibfield  {author} {\bibinfo {author} {\bibfnamefont {W.}~\bibnamefont
  {Paul}},\ }\href@noop {} {\bibfield  {journal} {\bibinfo  {journal} {Rev.
  Mod. Phys.},\ }\textbf {\bibinfo {volume} {62}},\ \bibinfo {pages} {531}
  (\bibinfo {year} {1990})}\BibitemShut {NoStop}%
\bibitem [{\citenamefont {Berkeland}\ \emph {et~al.}(1998)\citenamefont
  {Berkeland}, \citenamefont {Miller}, \citenamefont {Bergquist}, \citenamefont
  {Itano},\ and\ \citenamefont {Wineland}}]{berkeland_minimization_1998-1}%
  \BibitemOpen
  \bibfield  {author} {\bibinfo {author} {\bibfnamefont {D.~J.}\ \bibnamefont
  {Berkeland}}, \bibinfo {author} {\bibfnamefont {J.~D.}\ \bibnamefont
  {Miller}}, \bibinfo {author} {\bibfnamefont {J.~C.}\ \bibnamefont
  {Bergquist}}, \bibinfo {author} {\bibfnamefont {W.~M.}\ \bibnamefont
  {Itano}}, \ and\ \bibinfo {author} {\bibfnamefont {D.~J.}\ \bibnamefont
  {Wineland}},\ }\href@noop {} {\bibfield  {journal} {\bibinfo  {journal} {J.
  Appl. Phys.},\ }\textbf {\bibinfo {volume} {83}},\ \bibinfo {pages} {5025}
  (\bibinfo {year} {1998})}\BibitemShut {NoStop}%
\bibitem [{\citenamefont {Schuster}\ \emph {et~al.}(2005)\citenamefont
  {Schuster}, \citenamefont {Wallraff}, \citenamefont {Blais}, \citenamefont
  {Frunzio}, \citenamefont {Huang}, \citenamefont {Majer}, \citenamefont
  {Girvin},\ and\ \citenamefont {Schoelkopf}}]{schuster_ac_2005}%
  \BibitemOpen
  \bibfield  {author} {\bibinfo {author} {\bibfnamefont {D.~I.}\ \bibnamefont
  {Schuster}}, \bibinfo {author} {\bibfnamefont {A.}~\bibnamefont {Wallraff}},
  \bibinfo {author} {\bibfnamefont {A.}~\bibnamefont {Blais}}, \bibinfo
  {author} {\bibfnamefont {L.}~\bibnamefont {Frunzio}}, \bibinfo {author}
  {\bibfnamefont {R.-S.}\ \bibnamefont {Huang}}, \bibinfo {author}
  {\bibfnamefont {J.}~\bibnamefont {Majer}}, \bibinfo {author} {\bibfnamefont
  {S.~M.}\ \bibnamefont {Girvin}}, \ and\ \bibinfo {author} {\bibfnamefont
  {R.~J.}\ \bibnamefont {Schoelkopf}},\ }\Doi {10.1103/PhysRevLett.94.123602}
  {\bibfield  {journal} {\bibinfo  {journal} {Phys. Rev. Lett.},\ }\textbf
  {\bibinfo {volume} {94}},\ \bibinfo {pages} {123602} (\bibinfo {year}
  {2005})}\BibitemShut {NoStop}%
\bibitem [{\citenamefont {Cederberg}\ \emph {et~al.}(2006)\citenamefont
  {Cederberg}, \citenamefont {Fortman}, \citenamefont {Porter}, \citenamefont
  {Etten}, \citenamefont {Feig}, \citenamefont {Bongard},\ and\ \citenamefont
  {Langer}}]{Cederberg06}%
  \BibitemOpen
  \bibfield  {author} {\bibinfo {author} {\bibfnamefont {J.}~\bibnamefont
  {Cederberg}}, \bibinfo {author} {\bibfnamefont {S.}~\bibnamefont {Fortman}},
  \bibinfo {author} {\bibfnamefont {B.}~\bibnamefont {Porter}}, \bibinfo
  {author} {\bibfnamefont {M.}~\bibnamefont {Etten}}, \bibinfo {author}
  {\bibfnamefont {M.}~\bibnamefont {Feig}}, \bibinfo {author} {\bibfnamefont
  {M.}~\bibnamefont {Bongard}}, \ and\ \bibinfo {author} {\bibfnamefont
  {L.}~\bibnamefont {Langer}},\ }\href@noop {} {\bibfield  {journal} {\bibinfo
  {journal} {J. Chem. Phys.},\ }\textbf {\bibinfo {volume} {124}},\ \bibinfo
  {eid} {244305} (\bibinfo {year} {2006})}\BibitemShut {NoStop}%
\bibitem [{\citenamefont {Nitz}\ \emph {et~al.}(1984)\citenamefont {Nitz},
  \citenamefont {Cederberg}, \citenamefont {Kotz}, \citenamefont {Hetzler},
  \citenamefont {Aakre},\ and\ \citenamefont {Walhout}}]{Nitz84}%
  \BibitemOpen
  \bibfield  {author} {\bibinfo {author} {\bibfnamefont {D.}~\bibnamefont
  {Nitz}}, \bibinfo {author} {\bibfnamefont {J.}~\bibnamefont {Cederberg}},
  \bibinfo {author} {\bibfnamefont {A.}~\bibnamefont {Kotz}}, \bibinfo {author}
  {\bibfnamefont {K.}~\bibnamefont {Hetzler}}, \bibinfo {author} {\bibfnamefont
  {T.}~\bibnamefont {Aakre}}, \ and\ \bibinfo {author} {\bibfnamefont
  {T.}~\bibnamefont {Walhout}},\ }\href@noop {} {\bibfield  {journal} {\bibinfo
   {journal} {J. Mol. Spectrosc.},\ }\textbf {\bibinfo {volume} {108}},\
  \bibinfo {pages} {6 } (\bibinfo {year} {1984})}\BibitemShut {NoStop}%
\bibitem [{\citenamefont {Weinreb}\ \emph {et~al.}(1988)\citenamefont
  {Weinreb}, \citenamefont {Pospieszalski},\ and\ \citenamefont
  {Norrod}}]{Cryogenic1988}%
  \BibitemOpen
  \bibfield  {author} {\bibinfo {author} {\bibfnamefont {S.}~\bibnamefont
  {Weinreb}}, \bibinfo {author} {\bibfnamefont {M.}~\bibnamefont
  {Pospieszalski}}, \ and\ \bibinfo {author} {\bibfnamefont {R.}~\bibnamefont
  {Norrod}},\ }\href@noop {} {\bibfield  {journal} {\bibinfo  {journal} {IEEE
  MTT/S International Microwave Symposium},\ \bibinfo {pages} {945}} (\bibinfo
  {year} {1988})}\BibitemShut {NoStop}%
\bibitem [{\citenamefont {Carmichael}\ \emph {et~al.}(1994)\citenamefont
  {Carmichael}, \citenamefont {Tian}, \citenamefont {Ren},\ and\ \citenamefont
  {Alsing}}]{Carmichael1994}%
  \BibitemOpen
  \bibfield  {author} {\bibinfo {author} {\bibfnamefont {H.~J.}\ \bibnamefont
  {Carmichael}}, \bibinfo {author} {\bibfnamefont {L.}~\bibnamefont {Tian}},
  \bibinfo {author} {\bibfnamefont {W.}~\bibnamefont {Ren}}, \ and\ \bibinfo
  {author} {\bibfnamefont {P.}~\bibnamefont {Alsing}},\ }in\ \href@noop {}
  {\emph {\bibinfo {booktitle} {Cavity Quantum Electrodynamics}}},\ \bibinfo
  {series and number} {Advances in atomic, molecular, and optical physics},\
  \bibinfo {editor} {edited by\ \bibinfo {editor} {\bibfnamefont {P.~R.}\
  \bibnamefont {Berman}}}\ (\bibinfo  {publisher} {Academic Press},\ \bibinfo
  {address} {Boston},\ \bibinfo {year} {1994})\ pp.\ \bibinfo {pages}
  {381--423}\BibitemShut {NoStop}%
\bibitem [{\citenamefont {Staanum}\ \emph {et~al.}(2010)\citenamefont
  {Staanum}, \citenamefont {H{\o}jbjerre}, \citenamefont {Skyt}, \citenamefont
  {Hansen},\ and\ \citenamefont {Drewsen}}]{Staanum10}%
  \BibitemOpen
  \bibfield  {author} {\bibinfo {author} {\bibfnamefont {P.~F.}\ \bibnamefont
  {Staanum}}, \bibinfo {author} {\bibfnamefont {K.}~\bibnamefont
  {H{\o}jbjerre}}, \bibinfo {author} {\bibfnamefont {P.~S.}\ \bibnamefont
  {Skyt}}, \bibinfo {author} {\bibfnamefont {A.~K.}\ \bibnamefont {Hansen}}, \
  and\ \bibinfo {author} {\bibfnamefont {M.}~\bibnamefont {Drewsen}},\
  }\href@noop {} {\bibfield  {journal} {\bibinfo  {journal} {Nature Phys.},\
  }\textbf {\bibinfo {volume} {6}},\ \bibinfo {pages} {271} (\bibinfo {year}
  {2010})}\BibitemShut {NoStop}%
\bibitem [{\citenamefont {Hudson}(2009)}]{Hudson09}%
  \BibitemOpen
  \bibfield  {author} {\bibinfo {author} {\bibfnamefont {E.~R.}\ \bibnamefont
  {Hudson}},\ }\href@noop {} {\bibfield  {journal} {\bibinfo  {journal} {Phys.
  Rev. A},\ }\textbf {\bibinfo {volume} {79}},\ \bibinfo {eid} {032716}
  (\bibinfo {year} {2009})}\BibitemShut {NoStop}%
\bibitem [{\citenamefont {Vogelius}\ \emph {et~al.}(2006)\citenamefont
  {Vogelius}, \citenamefont {Madsen},\ and\ \citenamefont
  {Drewsen}}]{vogelius_rotational_2006}%
  \BibitemOpen
  \bibfield  {author} {\bibinfo {author} {\bibfnamefont {I.~S.}\ \bibnamefont
  {Vogelius}}, \bibinfo {author} {\bibfnamefont {L.~B.}\ \bibnamefont
  {Madsen}}, \ and\ \bibinfo {author} {\bibfnamefont {M.}~\bibnamefont
  {Drewsen}},\ }\Doi {10.1088/0953-4075/39/19/S32} {\bibfield  {journal}
  {\bibinfo  {journal} {J. Phys. B},\ }\textbf {\bibinfo {volume} {39}},\
  \bibinfo {pages} {S1267} (\bibinfo {year} {2006})}\BibitemShut {NoStop}%
\bibitem [{\citenamefont {Labaziewicz}\ \emph {et~al.}(2008)\citenamefont
  {Labaziewicz}, \citenamefont {Ge}, \citenamefont {Antohi}, \citenamefont
  {Leibrandt}, \citenamefont {Brown},\ and\ \citenamefont
  {Chuang}}]{labaziewicz_suppression_2008}%
  \BibitemOpen
  \bibfield  {author} {\bibinfo {author} {\bibfnamefont {J.}~\bibnamefont
  {Labaziewicz}}, \bibinfo {author} {\bibfnamefont {Y.~F.}\ \bibnamefont {Ge}},
  \bibinfo {author} {\bibfnamefont {P.}~\bibnamefont {Antohi}}, \bibinfo
  {author} {\bibfnamefont {D.}~\bibnamefont {Leibrandt}}, \bibinfo {author}
  {\bibfnamefont {K.~R.}\ \bibnamefont {Brown}}, \ and\ \bibinfo {author}
  {\bibfnamefont {I.~L.}\ \bibnamefont {Chuang}},\ }\href@noop {} {\bibfield
  {journal} {\bibinfo  {journal} {Phys. Rev. Lett.},\ }\textbf {\bibinfo
  {volume} {100}},\ \bibinfo {pages} {013001} (\bibinfo {year}
  {2008})}\BibitemShut {NoStop}%
\bibitem [{\citenamefont {Schmidt}\ \emph {et~al.}(2005)\citenamefont
  {Schmidt}, \citenamefont {Rosenband}, \citenamefont {Langer}, \citenamefont
  {Itano}, \citenamefont {Bergquist},\ and\ \citenamefont
  {Wineland}}]{schmidt_spectroscopy_2005}%
  \BibitemOpen
  \bibfield  {author} {\bibinfo {author} {\bibfnamefont {P.~O.}\ \bibnamefont
  {Schmidt}}, \bibinfo {author} {\bibfnamefont {T.}~\bibnamefont {Rosenband}},
  \bibinfo {author} {\bibfnamefont {C.}~\bibnamefont {Langer}}, \bibinfo
  {author} {\bibfnamefont {W.~M.}\ \bibnamefont {Itano}}, \bibinfo {author}
  {\bibfnamefont {J.~C.}\ \bibnamefont {Bergquist}}, \ and\ \bibinfo {author}
  {\bibfnamefont {D.~J.}\ \bibnamefont {Wineland}},\ }\href@noop {} {\bibfield
  {journal} {\bibinfo  {journal} {Science},\ }\textbf {\bibinfo {volume}
  {309}},\ \bibinfo {pages} {749} (\bibinfo {year} {2005})}\BibitemShut
  {NoStop}%
\bibitem [{\citenamefont {Blais}\ \emph {et~al.}(2007)\citenamefont {Blais},
  \citenamefont {Gambetta}, \citenamefont {Wallraff}, \citenamefont {Schuster},
  \citenamefont {Girvin}, \citenamefont {Devoret},\ and\ \citenamefont
  {Schoelkopf}}]{blais_quantum-information_2007}%
  \BibitemOpen
  \bibfield  {author} {\bibinfo {author} {\bibfnamefont {A.}~\bibnamefont
  {Blais}}, \bibinfo {author} {\bibfnamefont {J.}~\bibnamefont {Gambetta}},
  \bibinfo {author} {\bibfnamefont {A.}~\bibnamefont {Wallraff}}, \bibinfo
  {author} {\bibfnamefont {D.~I.}\ \bibnamefont {Schuster}}, \bibinfo {author}
  {\bibfnamefont {S.~M.}\ \bibnamefont {Girvin}}, \bibinfo {author}
  {\bibfnamefont {M.~H.}\ \bibnamefont {Devoret}}, \ and\ \bibinfo {author}
  {\bibfnamefont {R.~J.}\ \bibnamefont {Schoelkopf}},\ }\href@noop {}
  {\bibfield  {journal} {\bibinfo  {journal} {Phys. Rev. A},\ }\textbf
  {\bibinfo {volume} {75}},\ \bibinfo {pages} {032329} (\bibinfo {year}
  {2007})}\BibitemShut {NoStop}%
\bibitem [{\citenamefont {Lin}\ \emph {et~al.}(2009)\citenamefont {Lin},
  \citenamefont {Zhu}, \citenamefont {Islam}, \citenamefont {Kim},
  \citenamefont {Chang}, \citenamefont {Korenblit}, \citenamefont {Monroe},\
  and\ \citenamefont {Duan}}]{lin_large-scale_2009}%
  \BibitemOpen
  \bibfield  {author} {\bibinfo {author} {\bibfnamefont {G.-D.}\ \bibnamefont
  {Lin}}, \bibinfo {author} {\bibfnamefont {S.-L.}\ \bibnamefont {Zhu}},
  \bibinfo {author} {\bibfnamefont {R.}~\bibnamefont {Islam}}, \bibinfo
  {author} {\bibfnamefont {K.}~\bibnamefont {Kim}}, \bibinfo {author}
  {\bibfnamefont {M.-S.}\ \bibnamefont {Chang}}, \bibinfo {author}
  {\bibfnamefont {S.}~\bibnamefont {Korenblit}}, \bibinfo {author}
  {\bibfnamefont {C.}~\bibnamefont {Monroe}}, \ and\ \bibinfo {author}
  {\bibfnamefont {L.-M.}\ \bibnamefont {Duan}},\ }\Doi
  {10.1209/0295-5075/86/60004} {\bibfield  {journal} {\bibinfo  {journal}
  {Euro. Phys. Lett.},\ }\textbf {\bibinfo {volume} {86}},\ \bibinfo {pages}
  {60004} (\bibinfo {year} {2009})}\BibitemShut {NoStop}%
\bibitem [{\citenamefont {Flambaum}\ and\ \citenamefont
  {Kozlov}(2007)}]{Flambaum07}%
  \BibitemOpen
  \bibfield  {author} {\bibinfo {author} {\bibfnamefont {V.~V.}\ \bibnamefont
  {Flambaum}}\ and\ \bibinfo {author} {\bibfnamefont {M.~G.}\ \bibnamefont
  {Kozlov}},\ }\href@noop {} {\bibfield  {journal} {\bibinfo  {journal} {Phys.
  Rev. Lett.},\ }\textbf {\bibinfo {volume} {99}},\ \bibinfo {eid} {150801}
  (\bibinfo {year} {2007})}\BibitemShut {NoStop}%
\end{thebibliography}

%

\end{document}